\documentclass[sigconf,10pt]{acmart}

\AtBeginDocument{%
  }

\setcopyright{none}
\renewcommand\footnotetextcopyrightpermission[1]{} 
\usepackage{balance}
\usepackage{xspace}
\usepackage[group-separator={,}, detect-weight]{siunitx}
\usepackage{algorithm}
\usepackage{algpseudocode}
\usepackage{subfigure}
\usepackage{caption}
\usepackage{amsmath}
\usepackage{multirow}
\usepackage{booktabs}
\usepackage{pifont}
\usepackage{graphicx}
\usepackage{pgfplots}
\usepackage{colortbl}
\usepackage{array}

\newcommand{\colorbar}[2]{
  \begin{tikzpicture}
    \fill[red!30] (0,0) rectangle (#1*0.2,0.3);
    \node[anchor=center, text width=1.8cm, align=center] at (1.1,0.15) {#2};
  \end{tikzpicture}
}

\newcommand{\SystemName}{{\it Argus}\xspace}
\newcommand{\ie}{{\it i.e.}\xspace}
\newcommand{\eg}{{\it e.g.}\xspace}
\newcommand{\etal}{{\it et al.}\xspace}
\newcommand\para[1]{{\vspace{3pt} \bf \noindent #1}}

\begin{document}

\title{\huge Argus: Multi-View Egocentric Human Mesh Reconstruction Based on Stripped-Down Wearable mmWave Add-on}


\author{Di Duan$^{1}$, Shengzhe Lyu$^{2}$, Mu Yuan$^{1}$, Hongfei Xue$^{3}$, Tianxing Li$^{4}$, Weitao Xu$^{2}$, Kaishun Wu$^{5\dagger}$, Guoliang Xing$^{1\dagger}$}

\thanks{$\dagger$ Guoliang Xing and Kaishun Wu are the corresponding authors.}

\email{diduan@cuhk.edu.hk, shengzhe.lyu@my.cityu.edu.hk, muyuan@cuhk.edu.hk}
\email{hongfei.xue@charlotte.edu, litianx2@msu.edu, weitaoxu@cityu.edu.hk}
\email{wuks@hkust-gz.edu.cn, glxing@cuhk.edu.hk}

\affiliation{
    \institution{$^1$The Chinese University of Hong Kong, $^2$City University of Hong Kong, $^3$University of North Carolina at Charlotte, $^4$Michigan State University, $^5$Hong Kong University of Science and Technology (GZ)}
    \country{}
    }

\renewcommand{\shortauthors}{Duan et al.}

\begin{abstract}
In this paper, we propose \SystemName, a wearable add-on system based on stripped-down (\ie, compact, lightweight, low-power, limited-capability) mmWave radars. It is the first to achieve egocentric human mesh reconstruction in a multi-view manner. Compared with conventional frontal-view mmWave sensing solutions, it addresses several pain points, such as restricted sensing range, occlusion, and the multipath effect caused by surroundings. To overcome the limited capabilities of the stripped-down mmWave radars (with only one transmit antenna and three receive antennas), we tackle three main challenges and propose a holistic solution, including tailored hardware design, sophisticated signal processing, and a deep neural network optimized for high-dimensional complex point clouds. Extensive evaluation shows that \SystemName achieves performance comparable to traditional solutions based on high-capability mmWave radars, with an average vertex error of \SI{6.5}{\centi\meter}, solely using stripped-down radars deployed in a multi-view configuration. It presents robustness and practicality across conditions, such as with unseen users and different host devices.
\end{abstract}

\maketitle
\pagestyle{plain}

\section{Introduction}
\label{sec:introduction}
\begin{figure}[t]
  \centering
  \includegraphics[width=\linewidth]{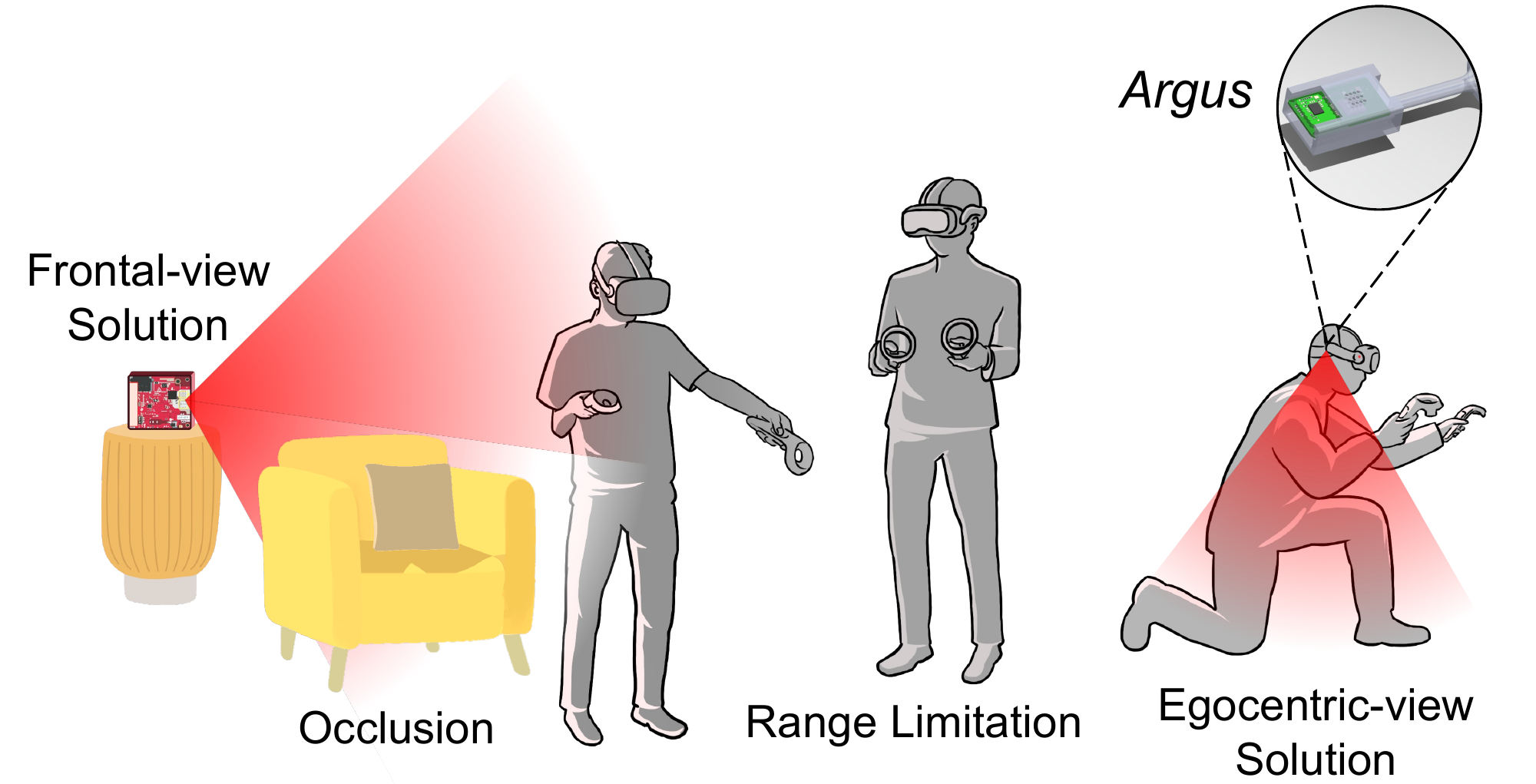}
  \caption{\SystemName is the first multi-view, egocentric mmWave sensing system enabling continuous HMR, breaking through the limitations of frontal-view solutions.}
  \label{fig:intro}
\end{figure}
Human pose estimation (HPE), including skeleton tracking and 3D human mesh reconstruction (HMR), remains a perennial research topic due to its broad application value in tasks such as fitness coaching~\cite{wang2019ai}, health monitoring~\cite{liu2022simultaneously}, and virtual reality~\cite{zhou2023metafi++}. Consequently, it has received significant attention from researchers~\cite{sun2022human}. Existing solutions attempted to solve this problem use various modalities, such as visual modalities~\cite{cao2017realtime, kocabas2020vibe, goel2023humans, kang2023ego3dpose, kang2024attention} (\eg, RGB, depth), wearable modalities~\cite{mollyn2023imuposer, devrio2023smartposer, yu2024seampose} (\eg, IMU, electromyography), and wireless modalities~\cite{zhao2019through, jiang2020towards, ren2021winect, xue2021mmmesh, lee2023hupr, shibata2023listening, mahmud2023posesonic} (\eg, ultrasound, Wi-Fi, mmWave). However, vision-based solutions are highly dependent on light conditions and struggle to darkness or smoke; while wearable-based solutions suffer from cumbersomeness and lack of user-friendliness. Based on this, many wireless solutions for HMR focus on using wireless signals. Among them, human sensing and reconstruction based on mmWave~\cite{zhang2023survey} is a representative research direction because it offers high precision, better penetration compared to Wi-Fi, and better interference resistance compared to low-frequency ultrasound. Based on this fact, it has garnered significant attention and led to many representative studies~\cite{xue2021mmmesh, chen2022mmbody, xue2022m4esh, zhang2022synthesized, xue2023towards, lee2023hupr, yang2024mmbat, zhang2024super} in the field.

Radio frequency (RF) signals are renowned for their non-contact, imperceptible, and user-friendly characteristics, making RF-based human sensing a subject of significant interest and leading to numerous practical applications~\cite{shuai2021millieye, xie2022mmfit, xu2022mmecg, qian20203d, dai2023interpersonal, han2024mmsign, bae2024supersight}. Previously, researchers successfully used Wi-Fi for human skeleton tracking~\cite{zhao2018rf} and mesh reconstruction~\cite{zhao2019through}. However, because of the ubiquity of Wi-Fi signals, they are prone to interference, and the sensing granularity is coarse. As a result, researchers have recently shifted towards using mmWave for human pose estimation and reconstruction, leading to the development of a series of studies. Xue \etal proposed mmMesh~\cite{xue2021mmmesh}, which is the first human mesh reconstruction solution based on a commodity mmWave radar. Later, several follow-up studies have been proposed to address remaining challenges or to improve the solution from different perspectives. For example, SynMotion~\cite{zhang2022synthesized} and mmGPE~\cite{xue2023towards} were later proposed to improve generalization by synthesizing mmWave signals; M$^4$esh~\cite{xue2022m4esh} and m$^3$Track~\cite{kong2022m3track} were proposed for multi-target tracking and reconstruction; several mmWave-native studies try to improve HPE performance by introducing an additional mmWave radar~\cite{lee2023hupr, zhang2024super} or employing advanced deep learning methods~\cite{yang2024mmbat}. However, all of the above works focus solely on mmWave-based HPE/HMR from a frontal view perspective, neglecting an intriguing perspective---\textit{mmWave-based HMR from an egocentric view}.

\begin{table*}
\caption{Comparison with other solutions (\ding{109}--Not Available,~\ding{108}--Available; Cons.: Consumption, TX \#: Number of Transmitters, RX \#: Number of Receivers).}
\resizebox{0.98\textwidth}{!}{%
\begin{tabular}{lcccccccccc}
\toprule
\textbf{Solutions} & \textbf{Radar} & \textbf{TX \#} & \textbf{RX \#} & \textbf{Board Size} & \textbf{Weight} & \textbf{Power Cons.} & \textbf{Multi-View} & \textbf{Sensing View} & \textbf{Body Part} \\ 
\hline
\addlinespace
{mmMesh~\cite{xue2021mmmesh}} & {AWR1843} & {3} & {4} & {\SI{8.3}{\centi\meter} $\times$ \SI{6.4}{\centi\meter}} & {$\sim$\SI{245}{\gram}} & {$\sim$\SI{2.08}{\watt}} & {\ding{109}} & {Frontal} & {Full} \\

{SynMotion~\cite{zhang2022synthesized}} & {IWR1443} & {3} & {4} & {\SI{8.3}{\centi\meter} $\times$ \SI{6.4}{\centi\meter}} & {$\sim$\SI{245}{\gram}} & {$\sim$\SI{2.1}{\watt}} & {\ding{109}} & {Frontal} & {Full} \\

{mmGPE~\cite{xue2023towards}} & {AWR1843} & {3} & {4} & {\SI{8.3}{\centi\meter} $\times$ \SI{6.4}{\centi\meter}} & {$\sim$\SI{245}{\gram}} & {$\sim$\SI{2.08}{\watt}} & {\ding{109}} & {Frontal} & {Full} \\

{M$^4$esh~\cite{xue2022m4esh}} & {AWR1843} & {3} & {4} & {\SI{8.3}{\centi\meter} $\times$ \SI{6.4}{\centi\meter}} & {$\sim$\SI{245}{\gram}} & {$\sim$\SI{2.08}{\watt}} & {\ding{109}} & {Frontal} & {Full} \\

{m$^3$Track~\cite{kong2022m3track}} & {AWR1443} & {3} & {4} & {\SI{7.8}{\centi\meter} $\times$ \SI{6.4}{\centi\meter}} & {$\sim$\SI{245}{\gram}} & {$\sim$\SI{2.1}{\watt}} & {\ding{109}} & {Frontal} & {Full} \\

{HUPR~\cite{lee2023hupr}} & {IWR1843} & {3} & {4} & {{\SI{7.8}{\centi\meter} $\times$ \SI{6.4}{\centi\meter}}} & {$\sim$\SI{245}{\gram}} & {$\sim$\SI{2.08}{\watt}} & {\ding{109}} & {Frontal} & {Full} \\

{SUPER~\cite{zhang2024super}} & {IWR6843} & {3} & {4} & {\SI{6.8}{\centi\meter} $\times$ \SI{5.5}{\centi\meter}} & {$\sim$\SI{137}{\gram}} & {$\sim$\SI{1.75}{\watt}} & {\ding{109}} & {Frontal} & {Upper} \\

{mmEgo~\cite{li2023egocentric}} & {IWR6843} & {3} & {4} & {\SI{6.8}{\centi\meter} $\times$ \SI{5.5}{\centi\meter}} & {$\sim$\SI{137}{\gram}} & {$\sim$\SI{1.75}{\watt}} & {\ding{109}} & {Egocentric} & {Full} \\ \midrule\midrule

{\textbf{Ours}} & {BGT60TR13C} & {1} & {3} & {\SI{3.9}{\centi\meter} $\times$ \SI{2.4}{\centi\meter}} & {$\sim$\SI{7.5}{\gram} $\times$ 2} & {$\sim$\SI{0.35}{\watt} $\times$ 2} & \ding{108} & {Egocentric} & {Full} \\ \bottomrule
\end{tabular}%
}
\label{tab:comparison}
\end{table*}
This interesting sensing view has several advantages over the frontal-view solutions. First, frontal-view solutions are subject to various limitations, such as restricted sensing range and susceptibility to interference from others’ movements (\ie, multipath effect) or occlusion. In contrast, using mmWave signals to sense a user from an egocentric perspective is highly promising to circumvent these challenges. Due to its on-body setup, the mmWave sensing field moves with the target, and the controllable sensing range can effectively avoid interference from others. Recently, the discovery of this ingenious research perspective brought about the first mmWave-based HMR solution from an egocentric view called mmEgo~\cite{li2023egocentric}. However, using such an egocentric view and mmWave signals to sense the user's human body typically implies a solution lies at the intersection of wearable and wireless sensing. Therefore, \textit{how to elegantly combine the characteristics of both remains a significant challenge}.

Although the mmWave radar (\ie, IWR6843ISK-ODS~\cite{tiIWR6843ISK-ODS}) used in mmEgo offers excellent sensing performance due to its multiple transmit and receive antennas, its size, weight, and high power consumption make it unsuitable for wearable solutions, rendering mmEgo far from practical. Motivated by this, we propose our solution---a pair of compact, lightweight, low-power mmWave-based add-ons named \SystemName that can be magnetically attached to various common host devices, such as VR headsets and headphones.

However, realizing this idea presents numerous unique challenges: \textbf{(1) How can multiple factors be fully considered in hardware design?} All previous studies are based on well-developed Texas Instruments (TI) mmWave radars, such as the IWR6843, IWR1843, IWR1443 series. These radars come with comprehensive development kits, including the mmWave SDK and mmWave Studio, which provide high-precision point cloud data, essential for advanced applications. However, these radars are relatively large in size and energy consumption, making them unsuitable for wearable devices that offer mmWave sensing in egocentric view. The development of \SystemName adopts compact sensors that help reduce the size of the hardware by significantly compromising the radar's capability, such as having fewer transmit and receive antennas. In addition, these compact radars often lack the sophisticated development kits needed to achieve high-precision point clouds, posing a major challenge of this paper. \textbf{(2) How to obtain the ground truth label in a more practical and user-friendly manner?} Previous solutions all rely on expensive and high-precision Motion Capture (MoCap) systems (\eg, VICON~\cite{vicon}, OptiTrack~\cite{optitrack}, Azure Kinect~\cite{azureKinect}) to provide high-quality labels for model training. However, such expensive MoCap systems are not commonly found in personal applications, and their deployment is often constrained by the available space, making mmWave-based HMR applications less widespread and challenging to implement practically. Therefore, acquiring high-quality labels for training using cost-effective commodity devices is a significant challenge in enhancing the practicality of the system. \textbf{(3) How does \SystemName overcome the dual challenges of self-occlusion and specular reflection?} Li \etal has already mentioned the issue of self-occlusion of the lower body by the upper body in egocentric views~\cite{li2023egocentric}. However, they used a tricky approach by deploying the mmWave radar extended far in front of the user’s head. Although this approach alleviates the self-occlusion problem and avoids specular reflections from the shoulders, it makes the system cumbersome and unsuitable for common small devices such as headphones. Therefore, effective integration of form factor design, signal processing pipeline, and deep neural network is essential to effectively address this intractable challenge and provide a better user experience.

To address these challenges, we proposed a series of solutions: \textbf{(1) To holistically consider the multiple factors inherent in hardware design}, we propose an innovative solution (Fig.~\ref{fig:intro}) that considers multiple factors by employing a pair of compact mmWave radars (\ie, BGT60TR13C~\cite{bgt60tr13c}) as an add-on, magnetically attached to the host device. \SystemName is the first portable and egocentric mmWave system for HMR that analyzes multi-view mmWave data (\ie, left and right), featuring a small size, lightweight, and low power consumption (details shown in Table~\ref{tab:comparison}). Furthermore, we have overcome the limitations imposed by the stripped-down mmWave radar through advanced signal processing techniques and deep learning. \textbf{(2) To facilitate the widespread application of \SystemName}, we employ monocular-based human mesh estimation using a single RGB-only camera (\eg, web camera, front camera of a smartphone) to acquire training labels, instead of relying on cumbersome and expensive MoCap systems. \textbf{(3) To overcome the challenges of self-occlusion and specular reflection}, we leverage multi-view sensing, namely, using dual egocentric-view mmWave sensing fields from left ear and right ear to collaboratively construct body meshes for the first time. Furthermore, we propose a tailored range-gating and energy-compensation solution for \SystemName. Moreover, Kolmogorov–Arnold Networks (KAN)~\cite{liu2024kan} is introduced to improve learning efficiency due to its superior capability in handling non-linearities, which are essential for modeling complex high-dimensional relationships (\eg, multi-view egocentric HMR). The contributions of this paper can be summarized as follows:
\begin{itemize}
    \item To the best of our knowledge, \SystemName is the first system realize the multi-view egocentric HMR by proposing a holistic solution, including prototype design, FMCW signal design and processing, and deep learning, etc.
    \item We first achieve multi-view mmWave sensing based on our wearable prototype. \SystemName is based on the joint analysis of mmWave fields with complementary fields of view, effectively addresses the self-occlusion and shoulder specular reflection issues.
    \item We propose and adopt a series of tailored techniques, such as clutter removal, range gating, energy compensation, and the introduction of KAN to enhance the system’s learning ability for high-dimensional non-linear representations. These techniques make it possible to reconstruct human meshes using compact and limited-capability mmWave radars.
    \item We perform a comprehensive evaluation of \SystemName, including its performance on unseen users and several micro-benchmarks. The evaluation results show that \SystemName outperforms two state-of-the-art (SOTA) baselines, demonstrating both robustness and practicality.
\end{itemize}
\section{Related Work}
\label{sec:related_work}
\para{mmWave-Based HPE and HMR.}
Driven by the contactless nature and high precision of mmWave sensing, HPE and HMR based on mmWave have received significant attention in recent years. Xue \etal proposed mmMesh~\cite{xue2021mmmesh}, using a commodity mmWave radar (\ie, AWR1843) for frontal-view HMR. However, there are still serveral challenges in this research area, such as the generalization to unseen activities, multi-target effect. Motivated by this, Xue \etal proposed their upgraded solutions M$^4$esh~\cite{xue2022m4esh} and mmGPE~\cite{xue2023towards}, which solve the problem of reconstructing multiple human meshes simultaneously and the generalization problem for unseen activities, respectively. To address the same problems in a different way, Zhang \etal proposed SynMotion~\cite{zhang2022synthesized} which synthesizes mmWave sensing signals to construct a mmWave dataset for generalization; m$^3$Track~\cite{kong2022m3track} is a contemporaneous work with M$^4$esh, it transforms the multi-target tracking task into a single-target HPE task. Furthermore, numerous studies have been proposed to improve performance by introducing multi-modality~\cite{chen2022mmbody}, extra mmWave radar from another sensing view (\ie, vertical and horizontal)~\cite{lee2023hupr, zhang2024super}, or more advanced deep learning techniques~\cite{wang2023human, yang2024mmbat}.

However, it seems that all the aforementioned works fall into a cognitive bias: \textit{assuming that the mmWave sensing for HPE or HMR must be performed from a frontal view}. In fact, not all studies have ignored egocentric-view sensing; mmEgo~\cite{li2023egocentric} was a pioneer in this idea. It prototyped a conventional mmWave radar (\ie, IWR6843) into a wearable device and made initial explorations of egocentric mmWave sensing. However, because of its size, weight, and power consumption, the design of mmEgo as a portable device was not usable, and the single-view sensing solution proved inadequate in addressing the challenges of self-occlusion and specular reflection. To fill this research gap, we propose \SystemName, the first multi-view sensing solution based on stripped-down mmWave radars. By introducing several innovative signal processing and deep learning techniques, it achieves acceptable performance despite significant hardware limitations, thereby improving the usability of egocentric HMR.

\begin{figure*}
  \centering
  \includegraphics[width=\linewidth]{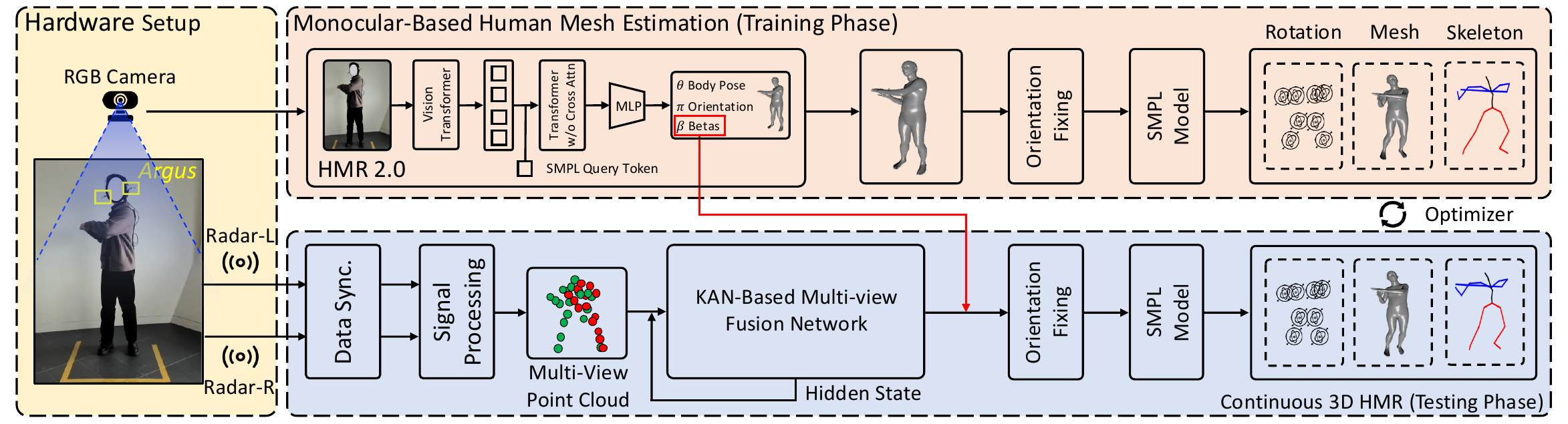}
  \caption{System overview. By transferring the knowledge from RGB images as pseudo-labels, a well-trained model will be deployed in the testing phase to perform continuous 3D human mesh reconstruction.}
  \label{fig:overview}
\end{figure*}
\para{Earable Sensing.}
Recently, mobile sensing based on earable devices has garnered widespread attention from researchers due to portability and informative sensing positions, which serve as a platform for multiple modalities, such as speech, ultrasound, electromyography (EMG), and photoplethysmogram (PPG). Numerous applications based on earable devices have been explored, such as speech enhancement~\cite{chatterjee2022clearbuds, veluri2023semantic, duan2024earse}, behavior recognition~\cite{prakash2020earsense, ma2021oesense, jiang2022earwalk, lyu2024earda, luo2024fuzzytrack}, health monitoring~\cite{bui2019ebp, chan2022off, butkow2023heart, chan2023wireless, hu2024breathpro, chen2024exploring}, and user authentication~\cite{fan2021headfi, wang2021eardynamic, wang2022toothsonic}. In addition, there have been security studies based on earable devices, covering both attack~\cite{liao2022magear} and defense~\cite{duan2024f2key}. A recent study, TinyssimoRadar~\cite{ronco2024tinyssimoradar}, successfully integrated mmWave modality with earable devices, achieving impressive performance in a simple gesture recognition task. However, the sensing capabilities of mmWave extend far beyond this. To the best of our knowledge, there has been no exploration of full-body pose estimation or mesh reconstruction using earable devices equipped with stripped-down mmWave radars. \SystemName is the first to tackle this challenging task under hardware limitations, filling this research gap.
\section{System Overview}
\label{sec:overview}
The system overview of \SystemName is shown in Fig.~\ref{fig:overview}, which contains two phases: a training phase in which the effective pseudo-labels obtained from RGB frames serve as the supervision to develop a KAN-based multi-view fusion network, and a testing phase where the well-trained deep neural network can continuously reconstruct multi-view mmWave frames into SMPL~\cite{pavlakos2019expressive} parameters.

Specifically, in the training phase, we collect visual stream data (RGB images) and mmWave stream data (data frames) using a commodity RGB camera and self-designed wearable add-ons (details in Sec.~\ref{subsec:hardware}), respectively. Then, we perform cross-modality data alignment using a two-tier method, which is elaborated in Sec.~\ref{subsec:synchronization}. Subsequently, the RGB frames are fed into the HMR 2.0~\cite{goel2023humans} neural network to estimate the SMPL parameters of these frames. Meanwhile, the corresponding mmWave data frames are processed by the proposed signal processing module (Sec.~\ref{subsec:signal_processing}) and further translated into the predicted SMPL parameters by the deep neural network (Sec.~\ref{subsec:network}). Since \SystemName is designed for on-body usage, an orientation-fixing module sets the matrix that represents the global orientation to the $3\times3$ identity matrix. Finally, the monocular-estimated and mmWave-predicted SMPL parameters are rendered into outputs (\ie, joint rotations, body meshes, skeletons) by the SMPL model. During training, the losses between these outputs are optimized together.

In the testing phase, the well-trained deep neural network will translate the processed multi-view mmWave features into SMPL parameters. The parameters, arranged as a time series, will be rendered into continuous body meshes or skeletons to support a corpus of applications such as e-fitness yoga instructor, avatar in the meta-universe, and virtual videoconferencing.
\section{System Design}
\label{sec:system_design}
\subsection{mmWave Stream \& Hardware Design}
\label{subsec:hardware}
As mentioned in Sec.~\ref{sec:introduction}, the hardware design of \SystemName should consider multiple factors in a comprehensive way, which causes specific difficulties that can be abstracted as follows:
\begin{itemize}
    \item The hardware prototype needs to support stable sensor configuration and data acquisition while maintaining a lightweight form factor as an add-on for common host devices, such as VR headsets and headphones.
    \item To facilitate effective sensor synchronization and efficient data transmission, a dedicated intermediary should be introduced between the local mobile devices and the two radar boards.
    \item It should be taken into account that user-friendliness and orientation-invariance of the device play a vital role in the performance and usability of the system.
\end{itemize}

\begin{figure}
    \centering
    \includegraphics[width=\linewidth]{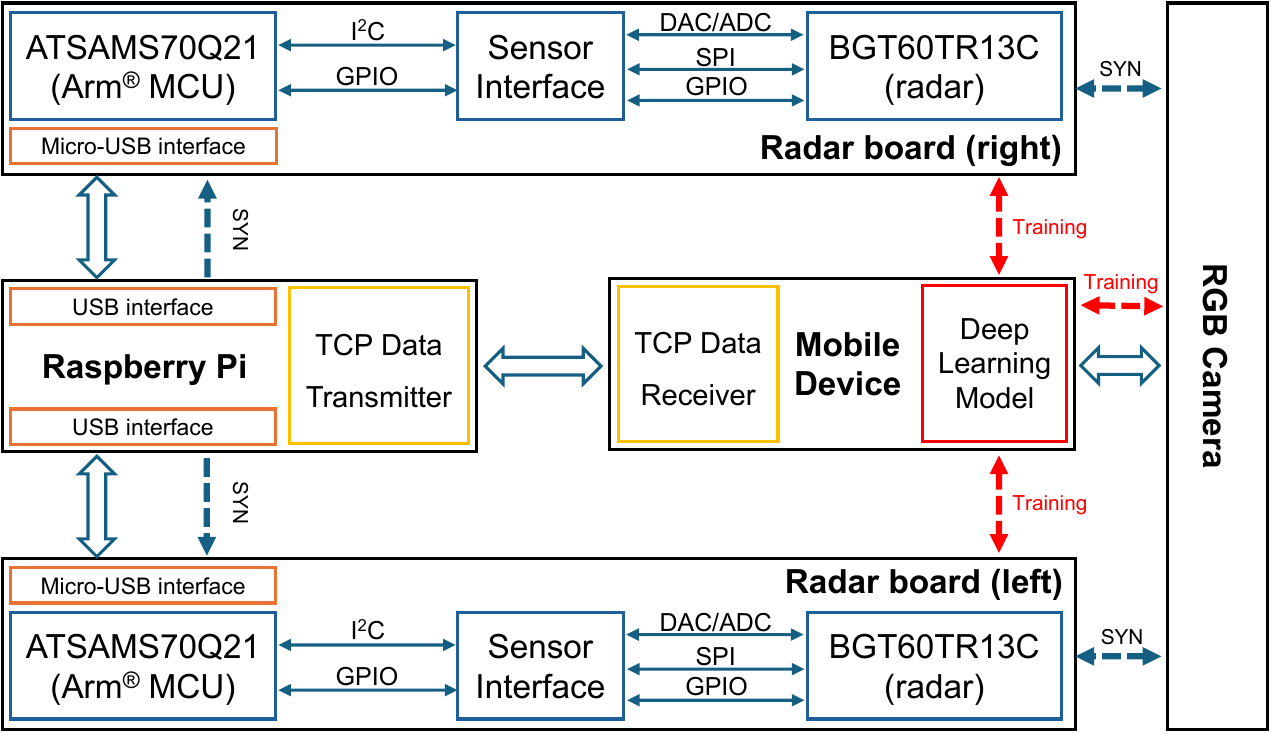}
    \caption{Block diagram of \SystemName hardware.}
    \label{fig:hardware_block}
\end{figure}

Taking into account the factors mentioned above, we designed the hardware shown in Fig.~\ref{fig:hardware_block} to implement our prototype. It consists of four main components: two radar boards for mmWave data acquisition, a Raspberry Pi for sensor synchronization and data transmission, and a mobile device for data receiving, processing, and rendering output results. Additionally, an RGB camera is utilized only during the training phase to obtain pseudo-labels for training. We now break down each of the components in detail.

\para{Radar boards.}
The radar board is designed to configure and transfer data from the mmWave sensor. It includes a baseboard with a Microchip ATSAMS70Q21 32-bit Arm Cortex-M7 MCU, and a daughter board hosting a BGT60TR13C~\cite{bgt60tr13c} 60 GHz mmWave radar sensor. The baseboard features two interfaces: a high-speed USB 2.0 connection for Raspberry Pi communication and a Serial Peripheral Interface (SPI) for mmWave sensor data transmission. It also integrates circuits for power management and debugging and the daughter board supports the radar chipset. The printed circuit board (PCB) measures $17 \times 12.7 \, \SI{}{\milli\meter\squared}$, with the mmWave sensor’s Antenna-in-Package (AIP) measuring $6.5 \times 5.0 \times 0.85 \, \SI{}{\milli\meter\cubed}$. The setup of the radar chip consists of one transmit antenna and three receive antennas arranged in an L-shape. The receiving antennas are grouped into two pairs, while the antennas are separated by half the wavelength (\ie, $\lambda/2$) in each pair. Low-pass filters are employed to minimize the impact of noise and crosstalk on the supply domains. Furthermore, an EEPROM connected via an I2C interface stores data such as the board identifier, while an \SI{80}{\mega \hertz} quartz oscillator ensures accurate timing for operations.

\para{Raspberry Pi.}
The synchronization and transmission center, built on a Raspberry Pi 4 with a Linux OS running specified radar Software Development Kit (SDK), acts as an intermediary between a mobile device and two radar boards. Powered by a power bank, the Raspberry Pi manages the synchronization, maintaining a time offset between the same frame from the left and right radars below \SI{10}{\milli\second}. It connects to the radar boards via USB interfaces to transmit configuration parameters and receive data. Wireless connectivity is facilitated by the built-in Wi-Fi module of the Raspberry Pi, allowing real-time communication with the mobile device. The multi-view mmWave data is transferred to the mobile device for computing in real time via a TCP protocol, ensuring the reliability and integrity of data transfer.

\begin{figure}
    \centering
    \begin{minipage}{0.4\linewidth}
        \centering
        \includegraphics[width=\linewidth]{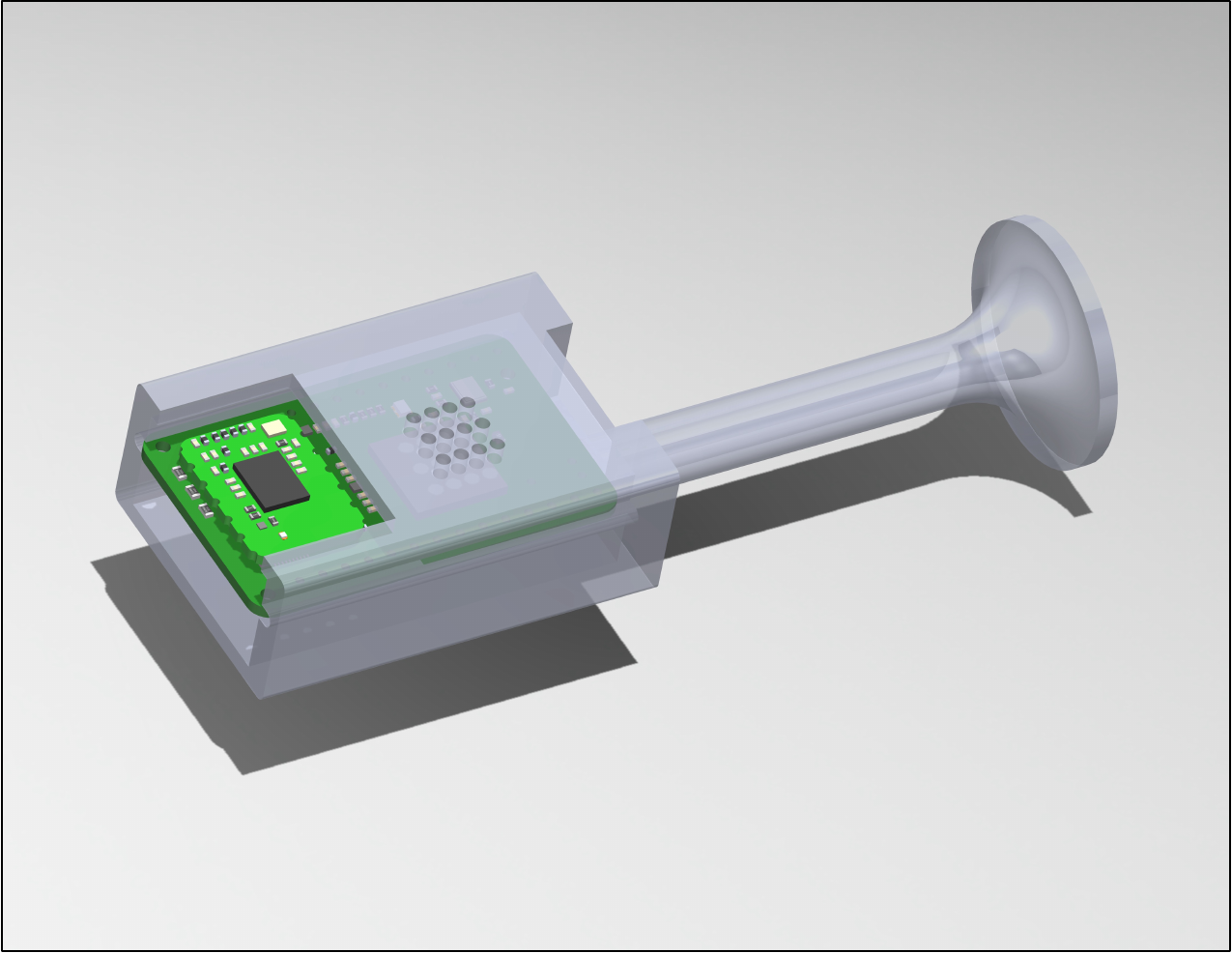}
        \caption{\SystemName 3D.}
        \label{fig:3d}
    \end{minipage}
    \hspace{1mm}
    \begin{minipage}{0.55\linewidth}
        \centering
        \includegraphics[width=\linewidth]{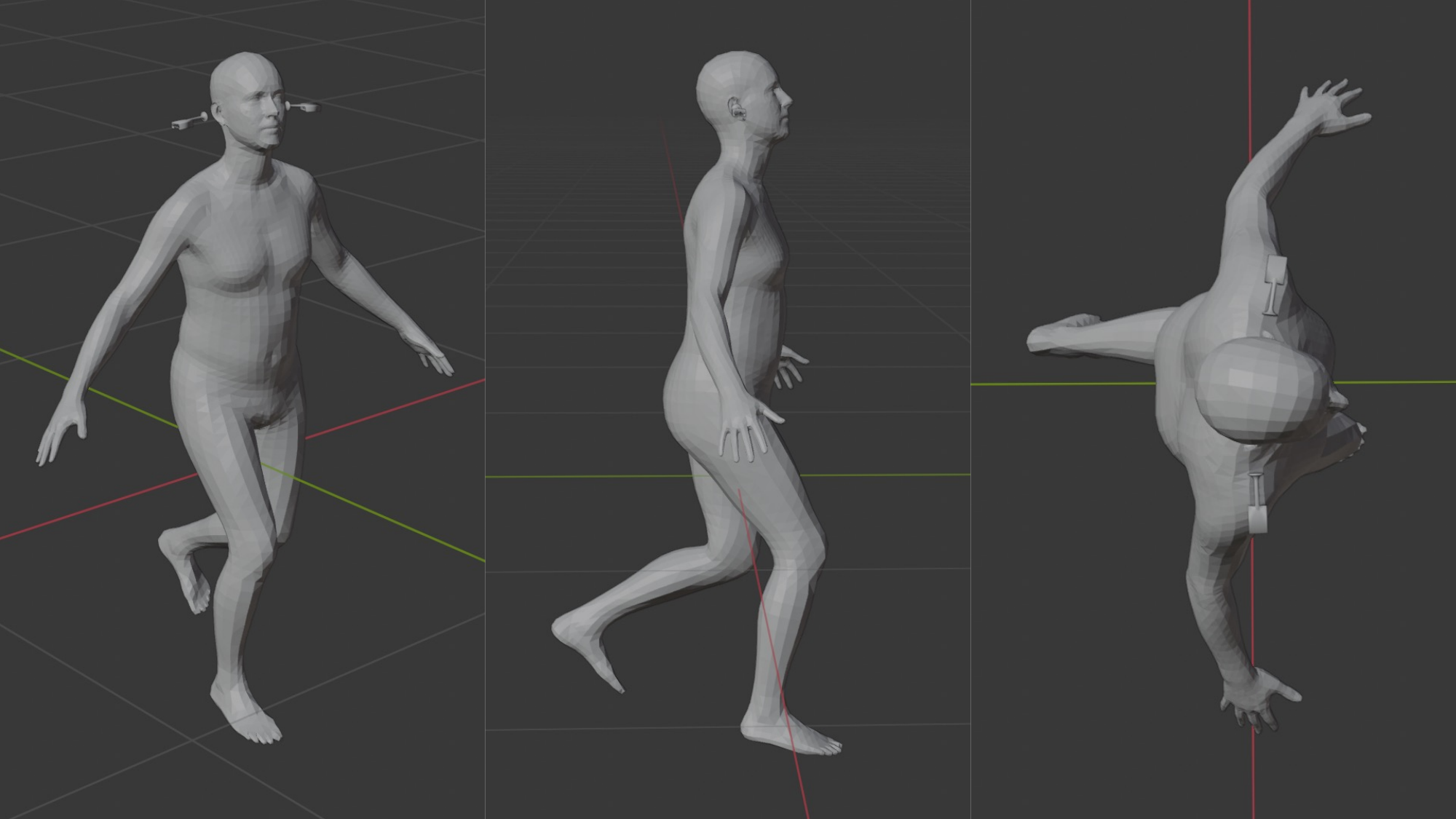}
        \caption{3D sensing model.}
        \label{fig:modeling}
    \end{minipage}
\end{figure}

\para{Mechanical Design and System Integration.}
The two radar boards are housed within 3D-printed enclosures (Fig.~\ref{fig:3d}), which are part of the integral \SystemName device designed to function as an add-on for general host devices (\eg, headphones). These transparent enclosures are custom-made for the radars of \SystemName, with one side featuring an open window for the emission of mmWave signals. Furthermore, the enclosures are designed to magnetically attach to host devices. As Fig.~\ref{subfig:detachable} shows, by carefully designing the pole arrangement of the four pairs of magnets on each side, it ensures that \SystemName add-ons are attached precisely to the same position each time and that left and right can be strictly distinguished. When the radar is placed on the wrong side, the pole arrangement creates repulsion rather than attraction. We modeled the sensing scenario in Blender~\cite{blender}, as shown in Fig.~\ref{fig:modeling}, where the add-ons are fixed to both sides of the user’s head, with the mmWave sensors positioned approximately \SI{8}{\centi\meter} horizontally from the user’s ears. Each mmWave sensor has a maximum field of view (FOV) of \SI{90}{\degree} and a maximum sensing distance of \SI{3.2}{\meter}. From the multiple perspectives of the simulated scenario shown in Fig.~\ref{fig:modeling}, it is easy to see that the sensing range of \SystemName can effectively cover and detect the movements of the user’s limbs.

In this section, we propose a holistic hardware solution that considers multiple factors such as stability, synchronization, and usability, addressing \textbf{the first key challenge} mentioned in Sec.~\ref{sec:introduction}.
\subsection{Visual Stream \& Pseudo-Label}
\label{subsec:pseudo_label}
Motivated by the high deployment overhead and the space limitation of cumbersome MoCap systems used in existing solutions~\cite{xue2021mmmesh, zhang2022synthesized, xue2022m4esh}, we pioneeringly propose a more practical approach to obtain pseudo-labels for training---\textit{estimating pose parameters from monocular images}.

In practice, we implement HMR 2.0, a cutting-edge human mesh estimation network based on monocular images, which includes a Vision Transformer (ViT) and a transformer decoder, on a commodity RGB camera. The ViT dissects the image into patches and processes these through self-attention mechanisms, allowing the model to capture global dependencies and intricate details across the entire image; while the cross-attention-based transformer decoder further refines the process by selectively focusing on relevant features extracted by the ViT. It dynamically adjusts its attention to specific image regions that are more informative in predicting human meshes, effectively dealing with occlusions and complex poses. By incorporating the ViT and transformer decoder, the network is designed to efficiently parse and understand the complexities of human poses and shapes from a monocular RGB image. The implementation of HMR 2.0 follows the open-source repository~\cite{HMR2_github}. However, to effectively use the pseudo-labels, we need to address another challenge---\textit{the accurate alignment between mmWave frames and RGB frames with different sampling rates}.
\subsection{Cross-Modality Data Alignment}
\label{subsec:synchronization}
\begin{algorithm}[t]
\caption{Cross-Modality Data Alignment}
\label{alg:align}
\begin{algorithmic}[1]
\small

\Require $N$: Number of frames, $\tau$: Threshold
\Ensure Data alignment within $\tau$ ms

\State $t_{I_{m}}, t_{L_{m}} \gets$ Query NTP, Record local time
\State $O_m \gets t_{I_{m}} - t_{L_{m}}$ \Comment{mmWave offset}

\State $t_{I_{i}}, t_{L_{i}} \gets$ Query NTP, Record local time
\State $O_i \gets t_{I_{i}} - t_{L_{i}}$ \Comment{RGB offset}

\For{$f = 1$ to $N$}
    \State ${t_{I_{m}}'}_f \gets {t_{L_{m}}}_f + O_m$ \Comment{Calibrate mmWave modality to NTP}
    \State ${t_{I_{i}}'}_f \gets {t_{L_{i}}}_f + O_i$ \Comment{Calibrate visual modality to NTP}
\EndFor

\For{each continuous sequence} \Comment{Across $N$ frames}
    \If{Avg $|t_{I_{m}}' - t_{I_{i}}'| > \tau$}
        \State Discard all $N$ frames
    \Else
        \State Use data for training
    \EndIf
\EndFor
\end{algorithmic}
\end{algorithm}
We propose two countermeasures to achieve the goal of data alignment, as shown in Algorithm~\ref{alg:align}. First, before capturing the mmWave data frames from both sides, the Raspberry Pi queries Network Time Protocol (NTP) servers using the ntplib library~\cite{ntplib} to obtain the NTP timestamp $t_{I_{m}}$ and records the local timestamp $t_{L_{m}}$ at the same time. By doing so, we can obtain the offset $O_{m}$ between the Raspberry Pi's local timestamps and the NTP timestamps, and further convert the local timestamps of the recorded mmWave frames into NTP timestamps. Similarly, we also calculate the time offset of the image modality that is represented as $O_{i}$. By converting the local timestamps of both modalities into NTP timestamps, the data from the two modalities can be aligned using the NTP timestamps. Second, during the model development phase, if the difference between the NTP timestamps of the mmWave data and the image data for a certain frame exceeds a threshold $\tau$, the data from that frame will be discarded and will not participate in the deep learning model training.

In scenarios involving time-series input, such as using a continuous sequence of N frames as a sample, it is necessary to evaluate whether the average timestamp difference across the N frames exceeds the threshold $\tau$. By implementing these two countermeasures together, the two modalities can be precisely aligned, and the synchronization error will be controlled within $\tau$. In this paper, we set $\tau$ to \SI{20}{\milli\second}. At this stage, by integrating the approaches from Sec.~\ref{subsec:pseudo_label} and Sec.~\ref{subsec:synchronization}, we address \textbf{the second key challenge} in Sec.~\ref{sec:introduction}.
\subsection{Signal Processing}
\label{subsec:signal_processing}
\begin{figure*}
  \centering
  \includegraphics[width=\linewidth]{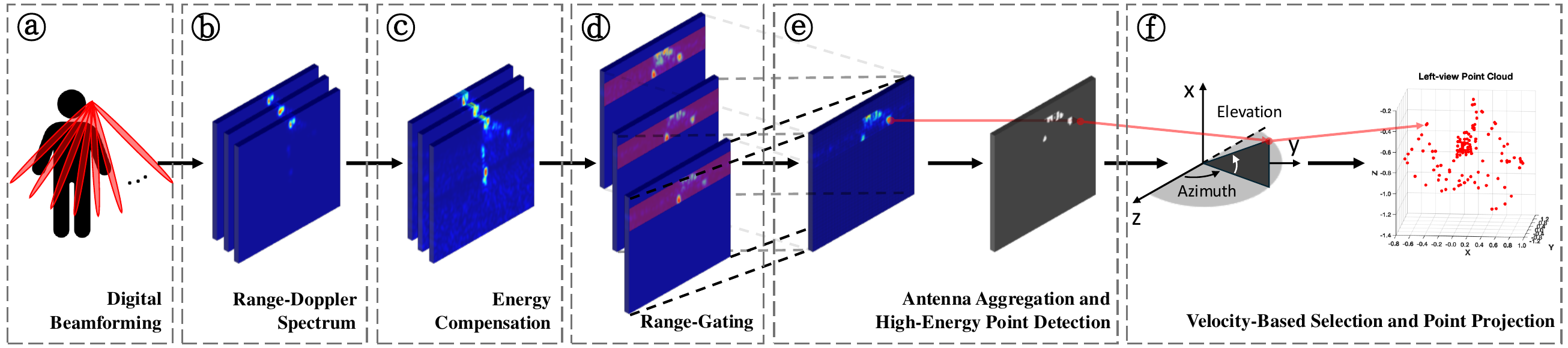}
  \caption{The pipeline for mmWave data signal processing.}
  \vspace{2mm}
  \label{fig:data_processing}
\end{figure*}
\begin{figure*}
    \centering
    \begin{minipage}{0.20\linewidth}
        \centering
        \includegraphics[width=\linewidth]{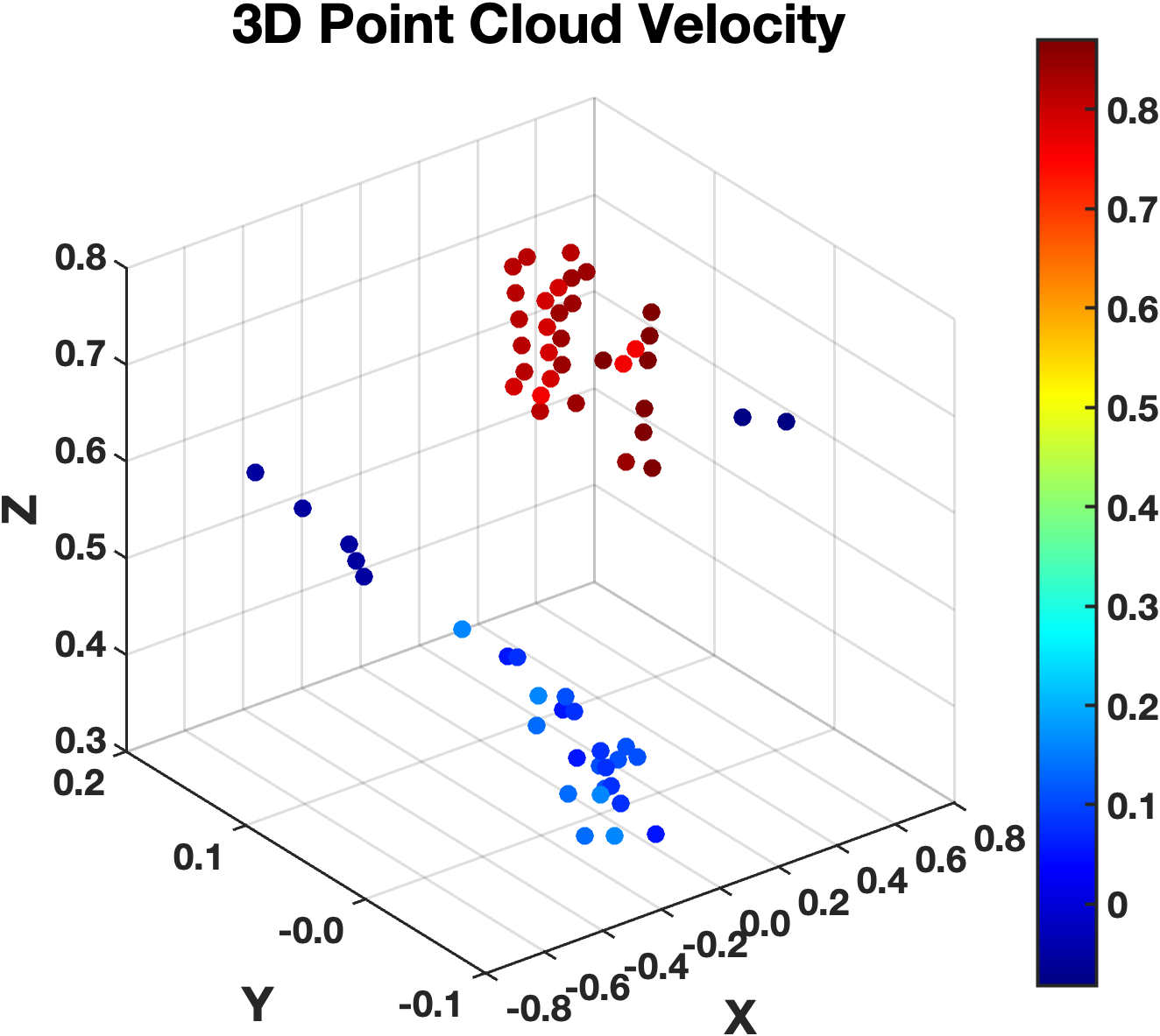}
        \caption{PC (Velocity).}
        \label{fig:pc_v}
    \end{minipage}
    \begin{minipage}{0.20\linewidth}
        \centering
        \includegraphics[width=\linewidth]{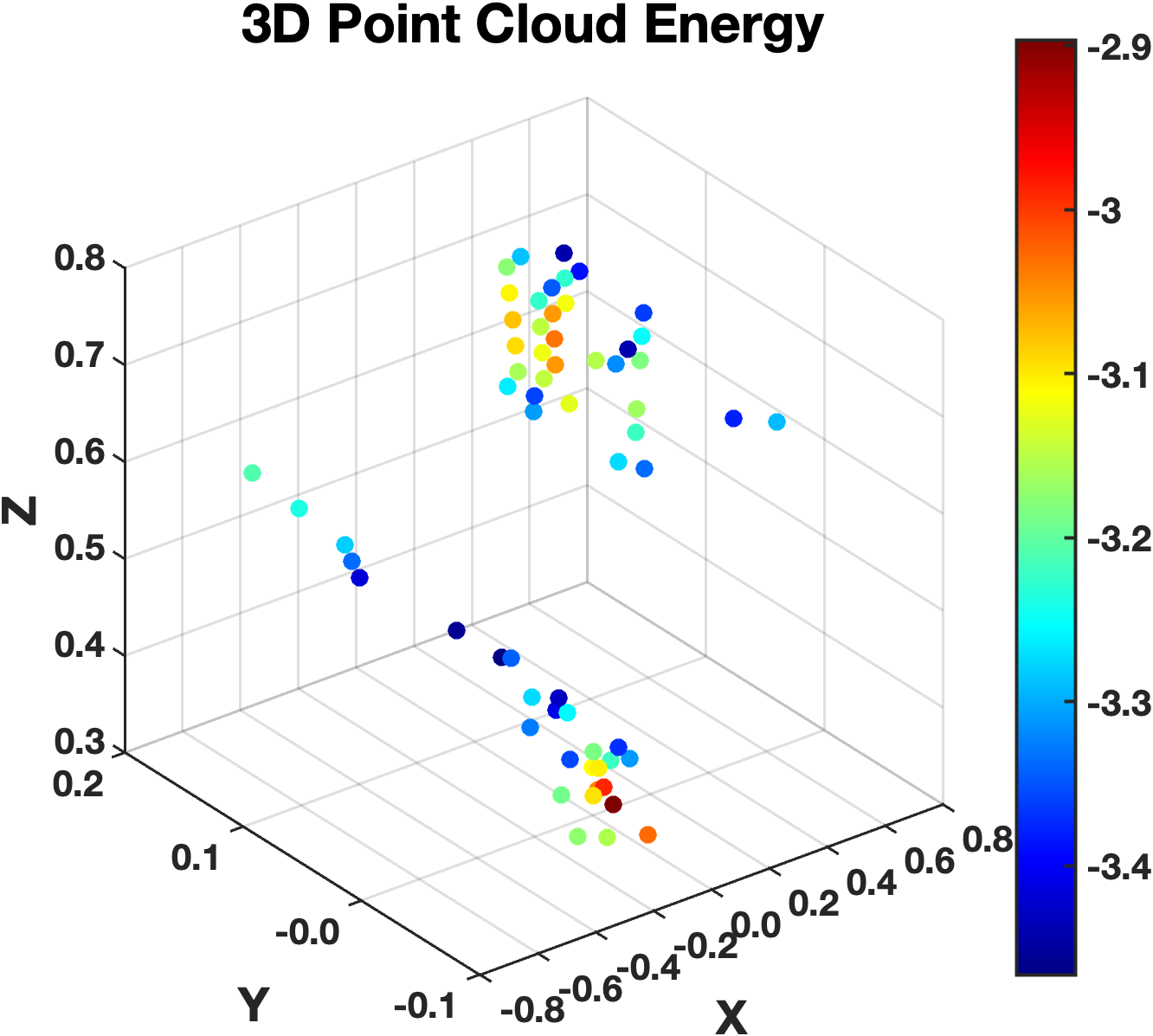}
        \caption{PC (Energy).}
        \label{fig:pc_e}
    \end{minipage}
    \begin{minipage}{0.20\linewidth}
        \centering
        \includegraphics[width=\linewidth]{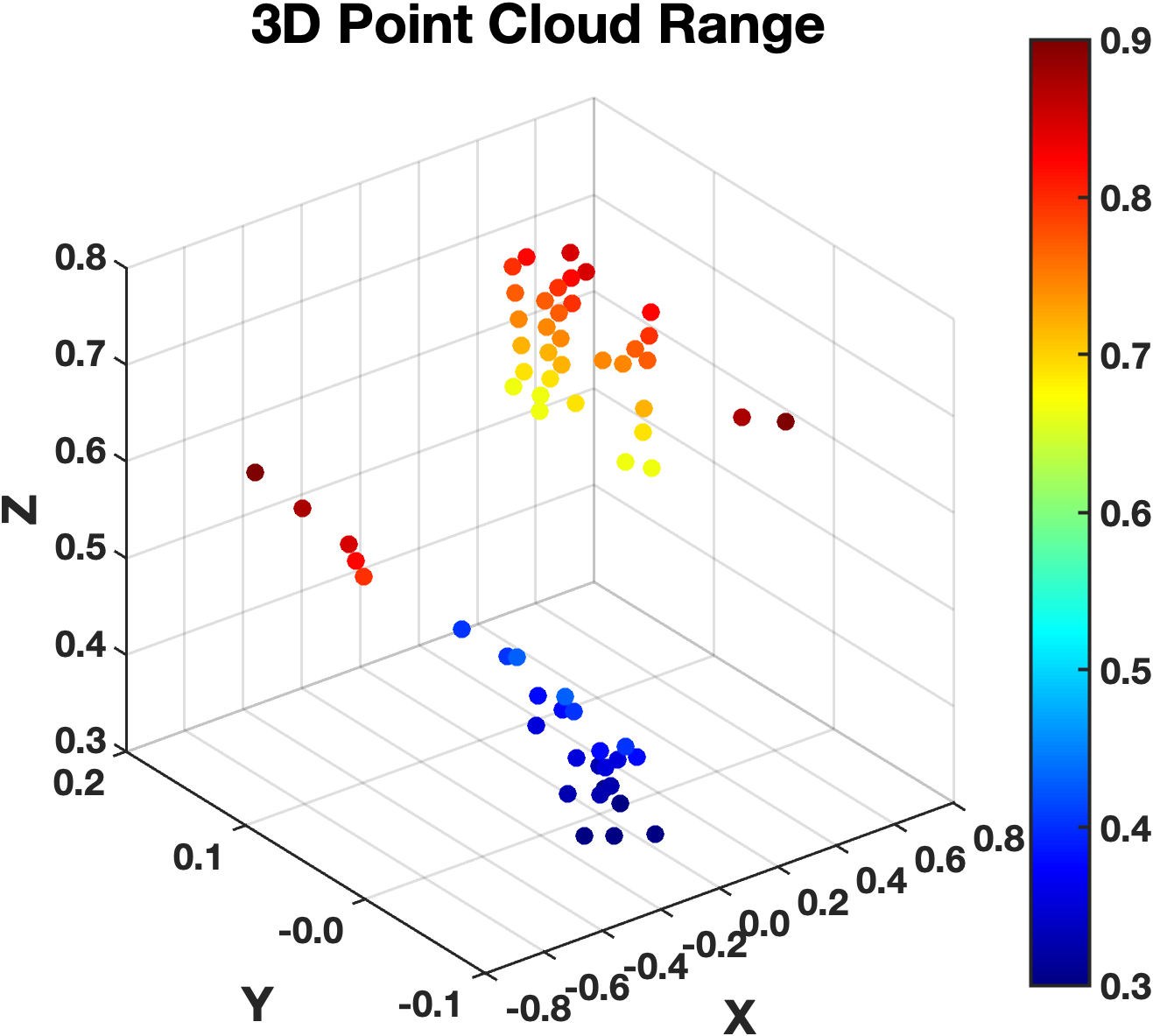}
        \caption{PC (Range).}
        \label{fig:pc_r}
    \end{minipage}
    \begin{minipage}{0.185\linewidth}
        \centering
        \includegraphics[width=\linewidth]{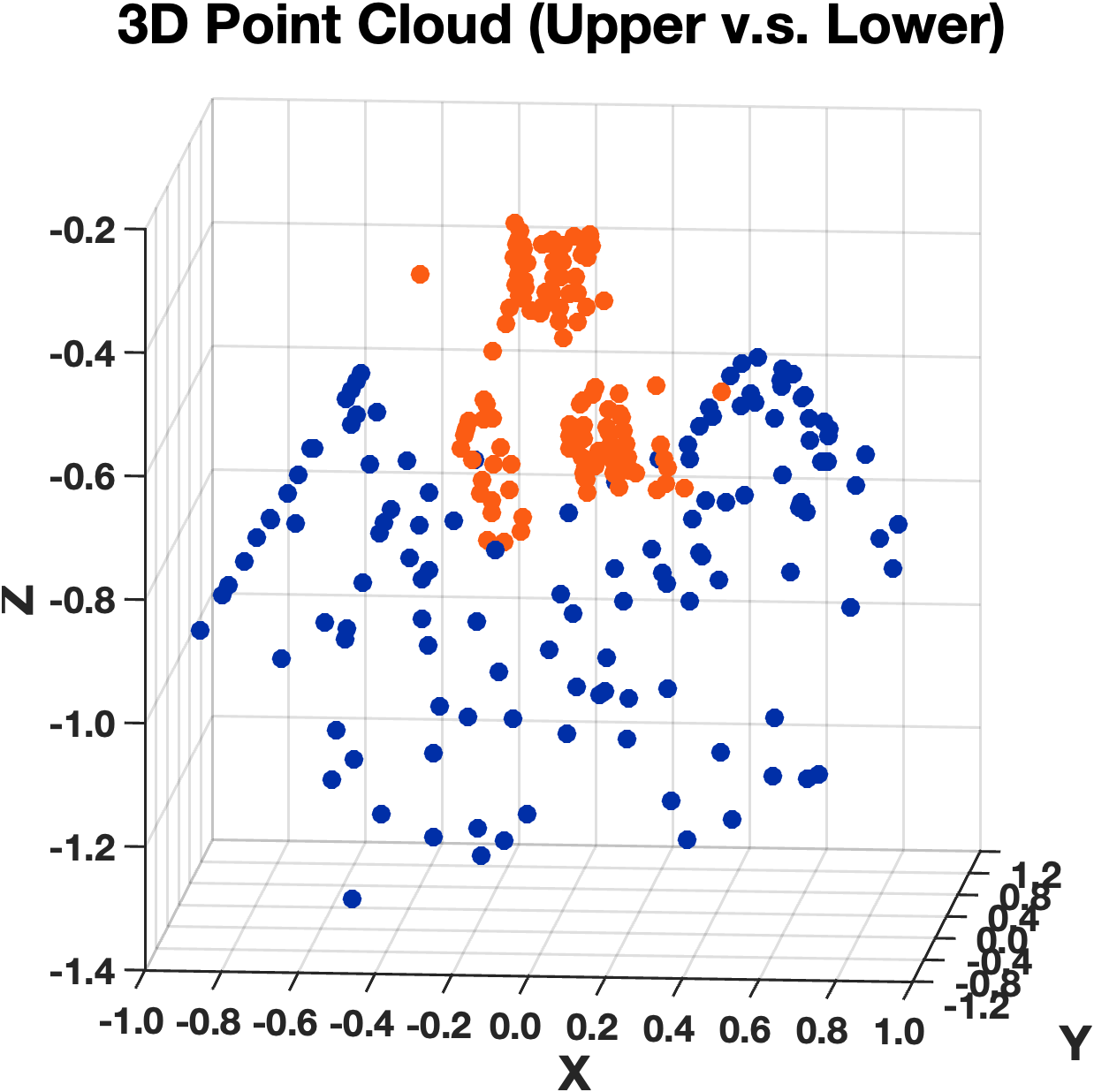}
        \caption{Range-gating.}
        \label{fig:pc_ul}
    \end{minipage}
    \begin{minipage}{0.185\linewidth}
        \centering
        \includegraphics[width=\linewidth]{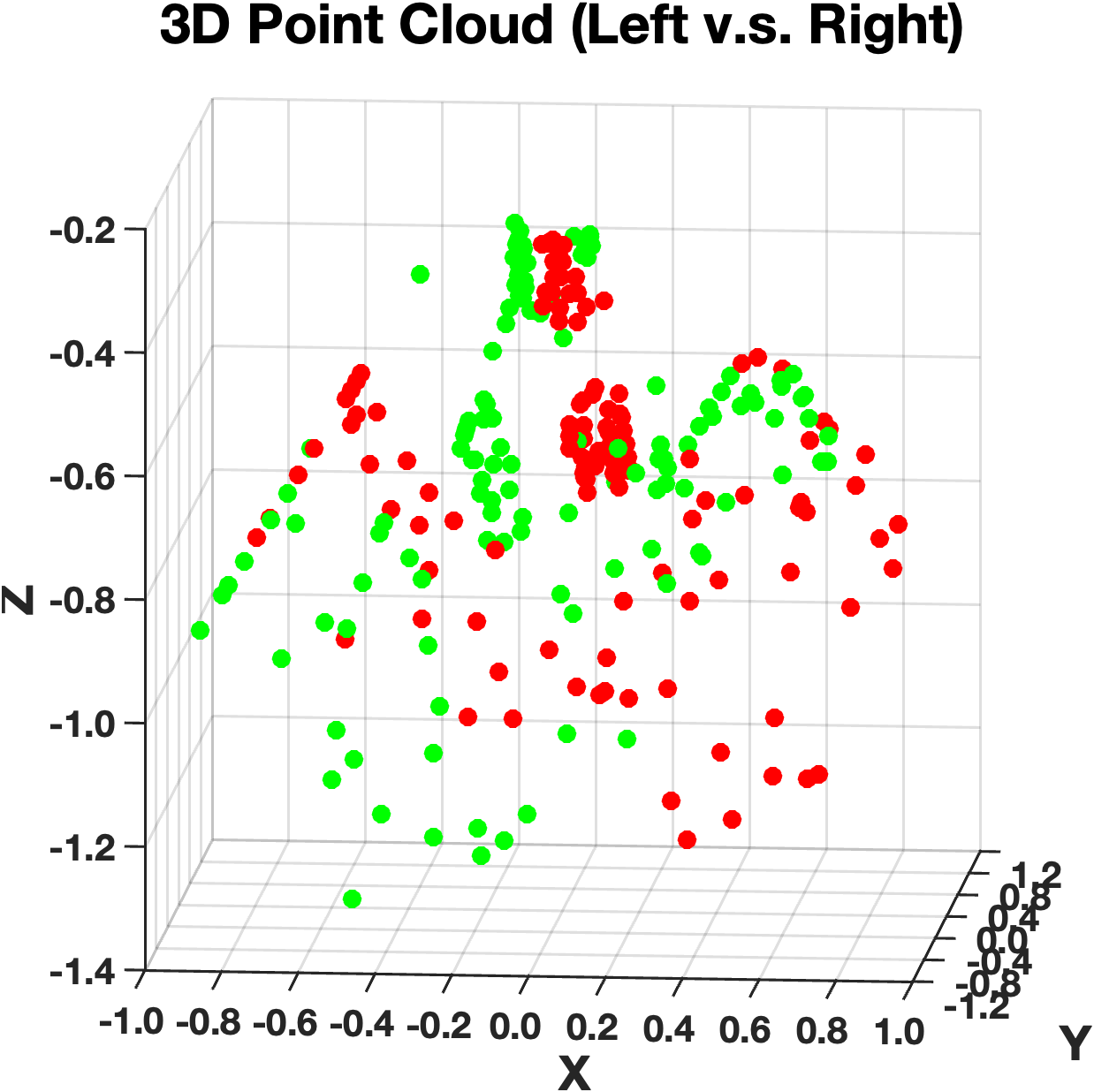}
        \caption{Multi-view.}
        \label{fig:pc_lr}
    \end{minipage}
\end{figure*}
To extract informative features for HMR from Frequency Modulated Continuous Wave (FMCW), we tailored a signal processing pipeline as illustrated in Fig.~\ref{fig:data_processing}. Specifically, the pipeline can be divided into six steps (\ie, \textcircled{a}--\textcircled{f}) and four main components: (1) Energy Compensation and Range-Gating for spatial attention; (2) Energy-Based and Velocity-Based Selection for informative points extraction; (3) Moving Target Indicator (MTI) and Clutter Removal for dynamic object attention; (4) Digital Beamforming (DBF) for striking a balance between the flexibility and structural regularity of the point cloud. Fig.~\ref{fig:pc_v}, Fig.~\ref{fig:pc_e}, and Fig.~\ref{fig:pc_r} present an example of the point cloud estimated from the signal processing pipeline. Since the range-Doppler maps used in \SystemName follow the standard FMCW signal processing procedure, we first briefly introduce our radar configuration and the range-Doppler maps, followed by an elaboration on the main components.

\begin{algorithm}[t]
\caption{Energy Compensation for Range-Doppler Map}
\label{alg:energy_compensation}
\begin{algorithmic}[1]
\small

\Require $rd_{s}$: Range-Doppler spectrum
\Ensure Compensated range-Doppler spectrum
\State $R, C \gets$ \text{shape}($rd_{s}$) \Comment{The number of range bins and channels}
\For{$c = 1$ to $C$}
    \State $M_{r} \gets$ \text{mean}(\text{abs}($rd_{s}[:, :, c]$, axis=1)) \Comment{Range bins energy}
    \State $M \gets$ \text{mean}($M_{r}$) \Comment{Mean energy across bins} 
    \For{$r = 1$ to $R$}
        \State $m \gets$ \text{mean}(\text{abs}($rd_{s}[r, :, c]$))
        \State $compensation \gets M / m$ \Comment{Scaling factor}
        \State $\tilde{rd_{s}}[r, :, c] \gets rd_{s}[r, :, c] \times compensation$
    \EndFor
\EndFor
\State \Return $\tilde{rd_{s}}$
\end{algorithmic}
\end{algorithm}
\para{Range-Doppler Maps.}
The radar configuration of \SystemName includes a frame period $T_f$ of \SI{100}{\milli\second}, containing $N_c$ of 128 chirps per frame. Each chirp is a linearly modulated continuous wave with a bandwidth $B$ of \SI{3}{\giga\hertz} (\SI{60}{}--\SI{63}{\giga\hertz}), consisting of $N_s$ samples and lasting $T_c$ time. In this paper, $N_s$ and $T_c$ are set to \SI{128}{} and \SI{700}{\micro\second}, respectively.

The transmitted signal reflects off objects and is received back, where the Intermediate Frequency (IF) signal's frequency, derived from mixing the received and transmitted signals, is given by: $f_{IF} = \frac{B}{T_c} \frac{2d}{c}$, where $d$ is the object distance. By stacking these IF signals across samples within each chirp and chirps within each frame, a radar frame sized $N_c \times N_s$ is constructed. Then, we apply the Fast Fourier Transform (FFT) twice to extract range and velocity information. Specifically, a \textit{Range FFT} along the sample axis of each radar frame results in 2D frames where the magnitudes signify the reflected energy of targets at different distances; a \textit{Doppler FFT} applied along the chirp axis yields range-Doppler maps, where the magnitude of each cell indicates the reflective energy of targets at specific ranges and velocities.

\para{Energy Compensation and Range-Gating.}
As discussed in Sec.~\ref{sec:introduction}, a significant challenge in egocentric mmWave sensing is the dual obstacle posed by self-occlusion and specular reflection. To address this challenge, it is essential to enhance the pipeline's spatial attention capability through signal processing techniques. To this end, we propose two ad hoc methods---Energy Compensation and Range-Gating.

Due to the energy attenuation of mmWave signals during transmission, the magnitudes representing the energy level in range-Doppler maps for the lower body are significantly weaker than those for the upper body. This further leads to the point cloud, estimated by detecting high-energy points from range-Doppler maps, being concentrated in the upper body. To mitigate the impact of energy attenuation, we designed an energy compensation method for range-Doppler maps, as shown in Algorithm~\ref{alg:energy_compensation}. We apply different scaling factors to range bins to make sure that the energy levels of them are the same after compensation. As depicted in step \textcircled{b} and \textcircled{c}, the proposed energy compensation increases the perception of the lower body range by improving spatial attention. Furthermore, as illustrated in step \textcircled{d}, we gate range-Doppler maps into two ranges (\ie, \SI{0.3}{\meter}--\SI{0.9}{\meter} and \SI{0.9}{\meter}--\SI{1.5}{\meter}) and estimate point clouds from each range, respectively. By doing so, it can eliminate the specular reflection from the shoulders and acromia (usually within \SI{0.3}{\meter}), and focus the lower body separately (Fig.~\ref{fig:pc_ul}).

\para{Energy-Based and Velocity-Based Selection.}
As Fig.~\ref{fig:data_processing} shows, we first detect the high-energy points over a threshold of \SI{-3.5}{\decibel} in the aggregated range-Doppler map (step \textcircled{e}) and project these points into a 3D space. Then, we filter the projected point cloud based on velocity (step \textcircled{f}), retaining only the top $N_{pn}$ points with the highest and lowest velocities; here we set $N_{pn}$ to 32 in this paper. Therefore, we can obtain 64 points from the upper/lower body in one view. Finally, we can obtain $64 \times 2 \text{ body parts} \times 2 \text{ views} = 256$ points from each mmWave frame. Using the energy-based and velocity-based selections above, the estimated point clouds provide valuable information related to motion. Fig.~\ref{fig:pc_lr} provides an example of the estimated point clouds from different views that complement each other, alleviating the problem of self-occlusion and enhancing the system’s sensing capability.

\para{Moving Target Indicator and Clutter Removal.}
Besides the signal processing steps visualized in Fig.~\ref{fig:data_processing}, we also focus on the removal of static objects, such as walls within the radar FOV, since they can interfere with the detection of moving targets and reduce the effectiveness of the radar.

\begin{figure}
    \centering
    \subfigure[Point cloud w/o MTI]{
        \includegraphics[width=0.45\linewidth]{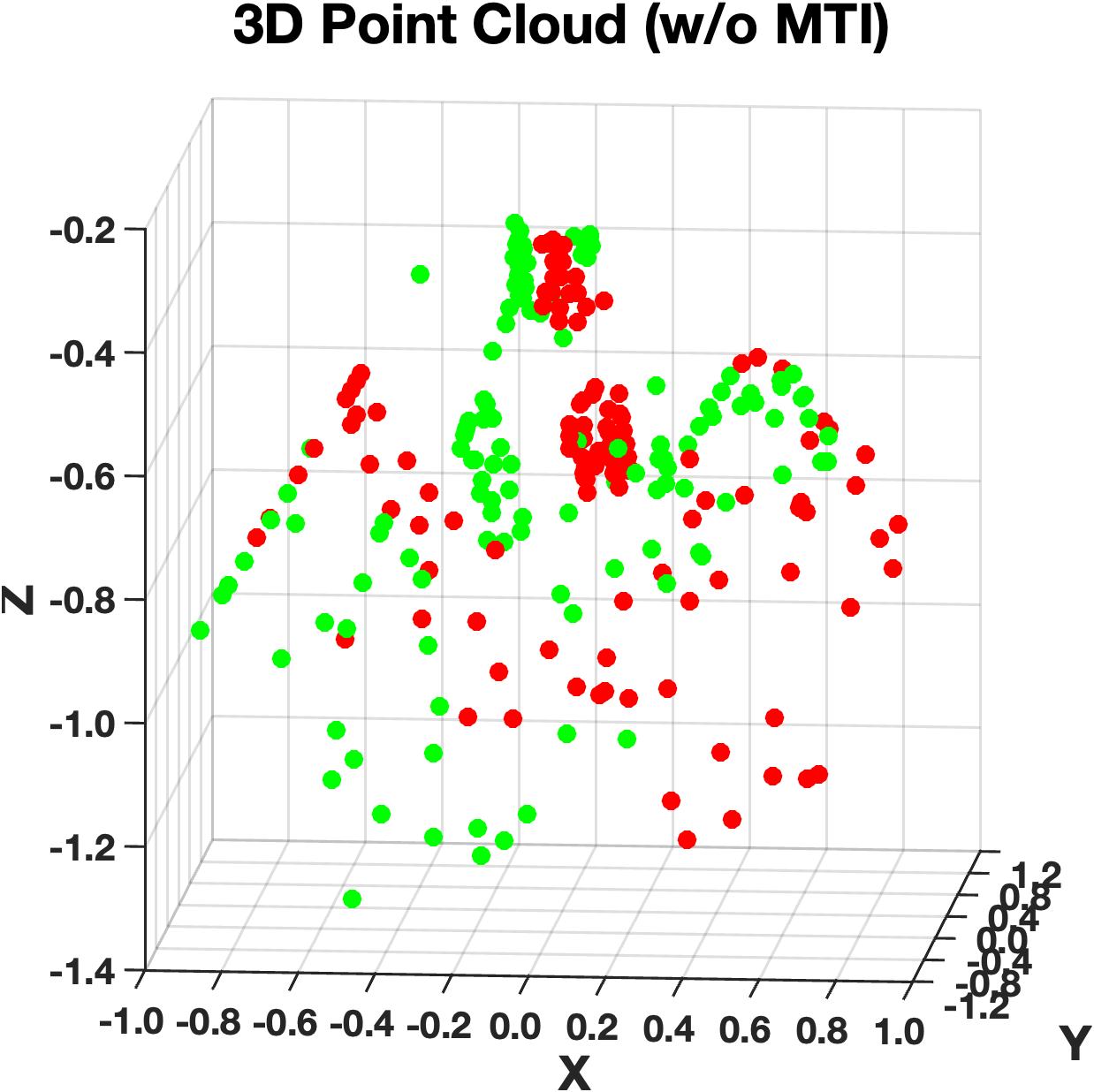}
        \label{subfig:woMTI}
    }
    \subfigure[Point cloud w/ MTI]{
        \includegraphics[width=0.45\linewidth]{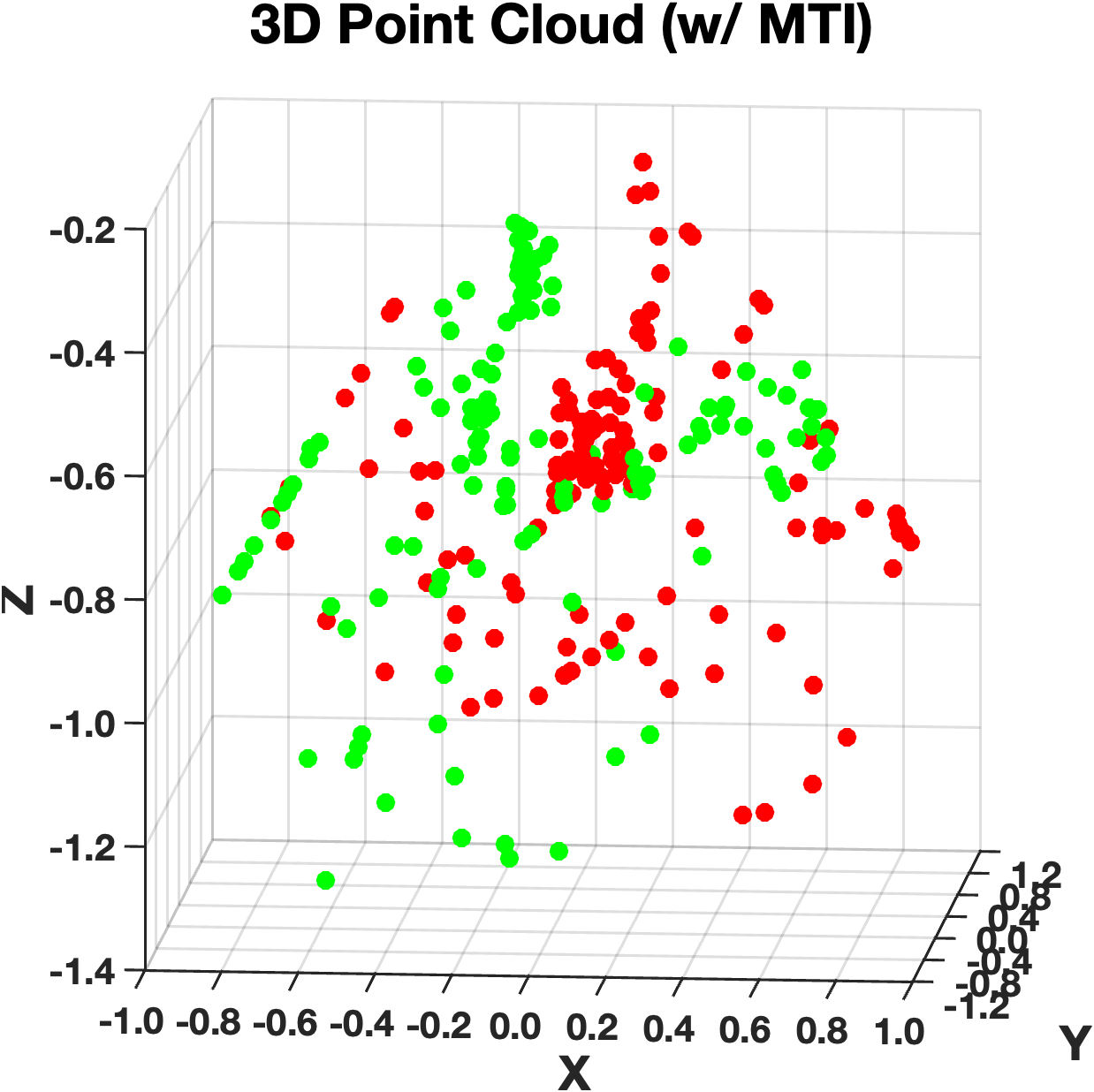}
        \label{subfig:wMTI}
    }
    \caption{Point clouds generated from the same mmWave frame with and without MTI.}
    \label{fig:MTI}
\end{figure}

To mitigate interference from static objects, the MTI~\cite{skolnik1980introduction} method is employed on raw data before Range FFT to suppress stationary clutter signals. Fig.~\ref{fig:MTI} shows the results of using MTI with the $\alpha_{MTI}$ set to 0.3 and the previous five mmWave frames for static object removal, before using MTI, the resulting point cloud is densely located in regions with smaller Z values ($\approx-0.3$) due to specular reflections from the shoulders and acromia. After using MTI, these less active regions are partially removed by considering them as static components. Since static components are effectively filtered out, employing a velocity-based selection method allows the selected fixed-count points to more accurately reflect the motion state of the target. Furthermore, \textit{clutter removal} is a critical step after performing the range FFT. By capturing the average signal level across all chirps at each range bin and subtracting this mean from the original input data, the impact of static components can be effectively removed, thereby enhancing the perception of human motions.

\begin{figure}
  \centering
  \includegraphics[width=0.95\linewidth]{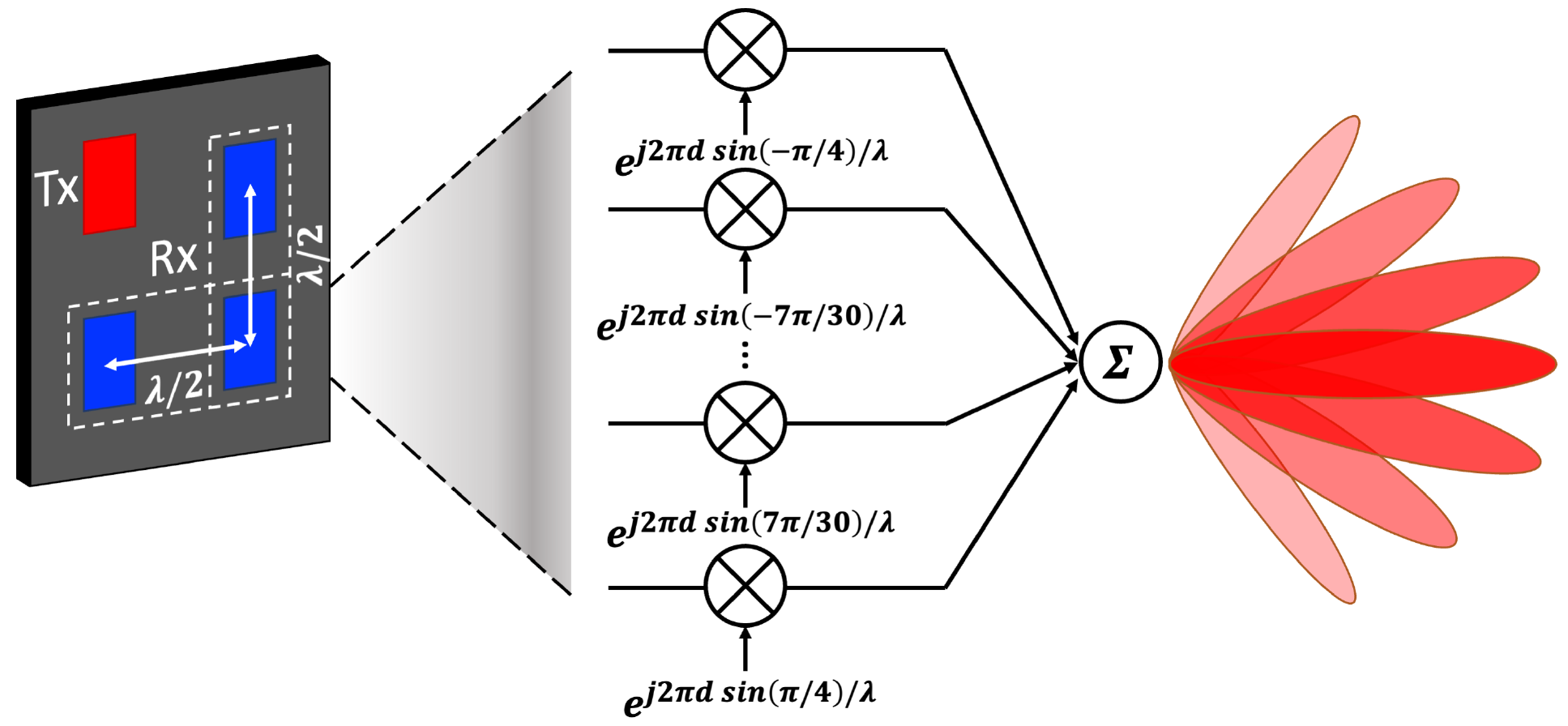}
  \caption{Digital beamforming uses a weight matrix to multiply the range-Doppler map of each antenna pair, forming beams corresponding to different directions.}
  \label{fig:DBF}
\end{figure}

\para{Digital Beamforming.}
We also apply DBF to reduce the randomness and dispersion of points by aligning them within defined spatial grids based on angle and range bins. This spatial alignment reduces ambiguity in point positions that could result from direct projection without DBF. Consequently, this process leads to more structured and consistent point clouds, which simplifies the learning process for deep models by providing cleaner and more reliable data for training. As Fig.~\ref{fig:DBF} shows, the three receiving antennas are grouped into two pairs, allowing the system to detect both the azimuth and elevation angles of an object relative to the radar. In DBF, the range-Doppler maps of two antennas in each pair are weighted individually by the weights matrix $W$ defined as:
\begin{equation*}
    \small
    W(i_{ant}, i_{beam}) = \exp\left( j \cdot 2\pi \cdot \frac{i_{ant}d}{\lambda} \cdot \sin(\theta_{beam}) \right),
\end{equation*}
where $W(i_{ant}, i_{beam})$ represents the weight for the $i_{ant}$ antenna and $i_{beam}$ beam, and $\theta_{beam}$ is the angle corresponding to the $i_{beam}$ beam in radians. For each beam, signals from different antennas are phase-shifted and combined. This constructive interference enhances the signal from the desired direction while suppressing signals from others. Therefore, the output of the weighting process represents the signal strength as a function of the arrival angle, calculated as:
\begin{equation*}
    \small
    \theta_{beam} = -\theta_{\text{max}} + \frac{2\theta_{\text{max}}}{N_{beam} - 1} \cdot i_{beam},
\end{equation*}
where $\theta_{\text{max}}$ is ${\pi}/4$ for this radar system, and the beam number $N_{beam}$ is set to 31 in this paper. As a result, the angle resolution of 3 degrees in 3D space is achieved. Compared to traditional angle estimation methods like Angle FFT, DBF not only balances flexibility and structural regularity but also allows for customizable $N_{beam}$, which achieves a balance between computational overhead and spatial resolution.
\subsection{Deep Neural Network Design}
\label{subsec:network}
\begin{figure*}
  \centering
  \includegraphics[width=0.95\linewidth]{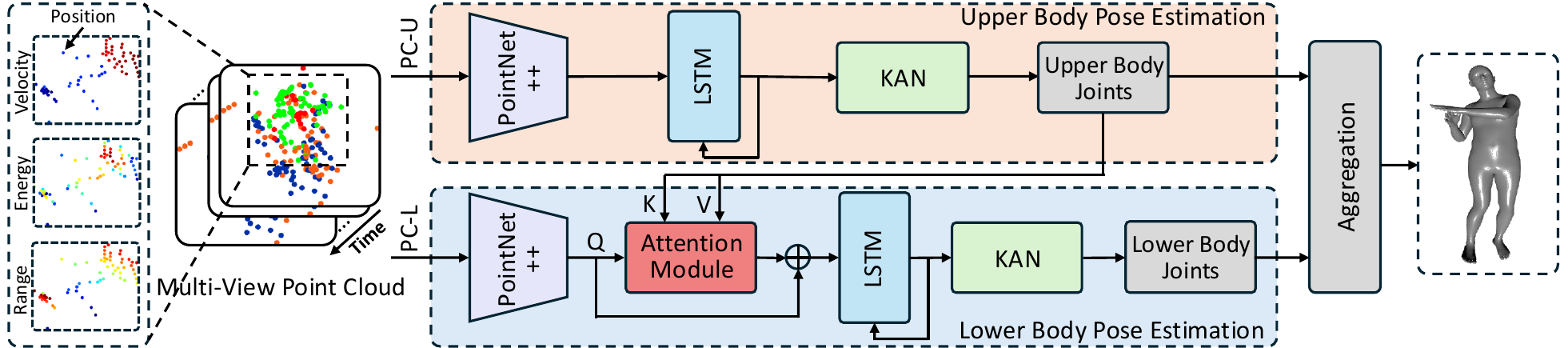}
  \caption{KAN-based multi-view fusion network.}
  \label{fig:network}
\end{figure*}

After we obtain multi-view point clouds from the upper and lower body separately, we elaborate on the design of the deep neural network used to translate multi-view point clouds into joint rotation matrices in this section. As Fig.~\ref{fig:network} shows, the neural network includes three main components: (1) PointNet++; (2) Long Short-Term Memory (LSTM) and Kolmogorov-Arnold Networks (KAN); (3) Multi-Head Attention. These components are designed to cooperate seamlessly to capture and process the complex spatial and temporal relationships present in the point clouds.

\para{Multi-Scale Point Cloud Feature Extractor.} For the specific task of HMR, it is crucial to extract point cloud features across multiple scales. The features extracted at different scales correspond to different levels of physical significance. As the scale radius increases from small to large, the extracted features transition from local to global, enabling more precise HMR. To achieve this goal, we use PointNet++~\cite{qi2017pointnet++} as the point cloud feature extractor to capture both local and global geometric features. It employs a series of Set Abstraction layers that progressively group points based on their spatial proximity and extract features from each group. This approach allows the model to learn multi-scale features effectively, making it well-suited for tasks that require detailed and hierarchical understanding of point cloud, such as HMR.

Specifically, we use three SA layers to progressively downsample the point cloud and extract multi-scale features. The first SA layer samples 128 points using radii of \SI{0.1}{\meter}, \SI{0.2}{\meter}, and \SI{0.4}{\meter}, with each radius having 16, 32, and 64 neighbors, respectively. The second SA layer further downsamples to 64 points with radii of \SI{0.2}{\meter}, \SI{0.4}{\meter}, and \SI{0.8}{\meter}, and the final SA layer aggregates global features without downsampling. Feature Propagation layers are then used to propagate the features back to the original point resolution, enabling the network to learn both local and global features effectively.

\para{LSTM and KAN.} Besides a powerful point cloud feature extractor, LSTM modules and KAN modules are also included in our deep neural network to serve as important roles.

Since \SystemName is designed to accept a series of sequential mmWave frames as inputs, we use the LSTM module to learn the relationships between adjacent frames to smooth the output. The hidden state from the previous time step (\ie, frame) in an LSTM is passed to the next time step and used, along with the current input, to compute the next hidden state. This allows the LSTM to capture and maintain temporal dependencies across the sequence, enabling the modeling of long-term dependencies. After obtaining the deep features of the current frame, we use KAN~\cite{liu2024kan} to transform these deep features into the dimensions required by the SMPL model. The core principle of KAN can be represented as follows:
\begin{equation*}
    \small
    f(x_1, x_2, \dots, x_n) = \sum_{j=1}^{2n+1} \Phi_j \left( \sum_{i=1}^n \phi_{ij}(x_i) \right),
\end{equation*}
where any multivariate continuous function $f(x_1, x_2, \dots, x_n)$ can be decomposed into a sum of compositions of univariate continuous functions $\phi_{ij}(x_i)$ and $\Phi_j$. It offers two significant advantages: (1) KAN, based on the Kolmogorov-Arnold representation theorem~\cite{kolmogorov1963representation, arnol1957functions}, can theoretically approximate any complex multivariate function more effectively. For tasks like HMR, which require high-precision multivariate function approximation, KAN provides a more refined function mapping capability with fewer parameters than fully connected layers, thereby improving reconstruction performance; (2) The structure of KAN allows for better decoupling of different input dimensions, which is particularly advantageous when handling high-dimensional point clouds (\eg, view, body part). This reduces model complexity, mitigates the risk of overfitting, and benefits HMR, where joint rotations involve highly non-linear and complex dependencies.

\para{Multi-Head Attention.} As mentioned in Sec.~\ref{sec:introduction}, a main challenge in achieving egocentric mmWave sensing is the problem of self-occlusion. To address this challenge, \SystemName features a hardware design that enables multi-view sensing, and we further fuse the information from the upper body to promote a more precise construction of the user’s lower body. Specifically, we use the prediction of the upper body pose $P_{u}$ as the K and V vectors in the attention module and feed the features of the lower body $F_{l}$ as the Q vector. This process can be formulated as:
\begin{equation*}
    \small
    \operatorname{MHA}(Q_{l}, K_{u}, V_{u}) = \operatorname{softmax}\left(\frac{F_{l} W_{Q_{l}} P_{u} W_{K_{u}}^{\top}}{\sqrt{d_{kl}}}\right) P_{u} W_{V_{u}},
\end{equation*}
where $W_{Q_{l}}$, $W_{K_{u}}$, and $W_{V_{u}}$ are the projection matrices for the Q, K, and V vectors, and $d_{kl}$ is the scaling factor. The reason behind this design is to leverage the global information encoded in the upper body prediction to guide the lower body pose estimation, ensuring consistency and coherence between the two. The attention mechanism can effectively query the relevant global features from the upper body, helping to refine and adjust the lower body prediction. This method preserves the hierarchy of information, where the prediction of the lower body is guided by the upper body, leading to more reasonable and precise pose estimation. Our deep neural network, with 42.4K parameters, is well-designed to run smoothly on mobile devices.

With this combination of the signal processing pipeline (Sec.~\ref{subsec:signal_processing}) and the multi-view fusion network (Sec.~\ref{subsec:network}), \textbf{the third key challenge} in Sec.~\ref{sec:introduction} is well addressed.
\begin{figure*}
    \centering
    \subfigure[Experimental scenario.]{
        \includegraphics[width=0.3\linewidth]{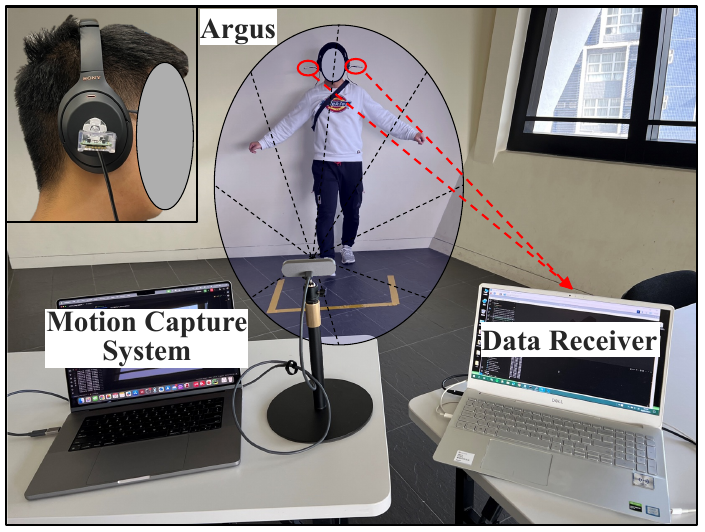}
        \label{subfig:scenario}
    }
    \hspace{-2mm}
    \subfigure[Wearing diagram.]{
        \includegraphics[width=0.174\linewidth]{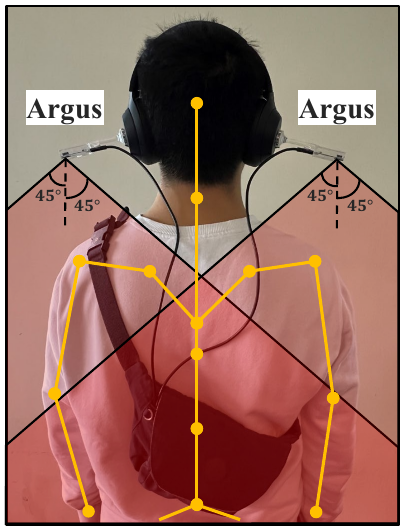}
        \label{subfig:wearing}
    }
    \hspace{-2mm}
    \subfigure[Details of \SystemName.]{
        \includegraphics[width=0.3\linewidth]{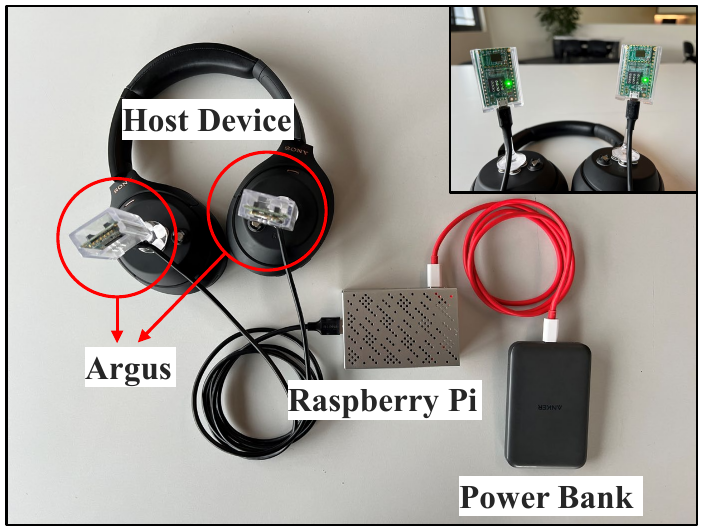}
        \label{subfig:details}
    }
    \hspace{-2mm}
    \subfigure[Detachable design.]{
        \includegraphics[width=0.174\linewidth]{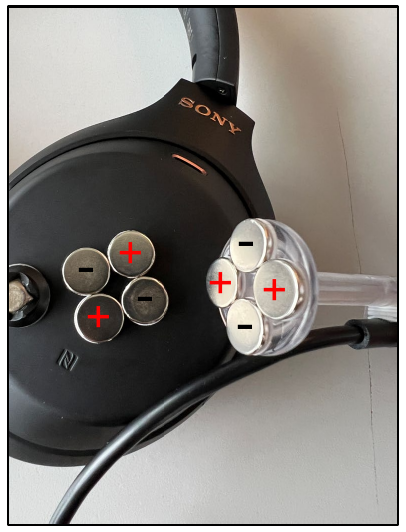}
        \label{subfig:detachable}
    }
    \caption{The testbed of \SystemName. The camera is used only to obtain pseudo-labels for training; once the model is trained and deployed on mobile devices, user movements are no longer restricted by it.}
    \label{fig:testbed}
\end{figure*}
\section{Evaluation}
\label{sec:evaluation}
\subsection{Testbed and Experimental Settings}
\label{subsec:testbed}
\para{Testbed.}
We use a monocular RGB camera (Logitech BRIO Ultra HD Pro~\cite{LogiBRIO4k}) combined with the advanced image-to-mesh approach (HMR 2.0~\cite{goel2023humans}) to extract pose parameters for the SMPL model from monocular images. Specifically, as Fig.~\ref{fig:testbed} shows, we place the camera in front of the participant and delineate a \SI{1}{\meter} $\times$ \SI{1}{\meter} area. Note that, participants' activities can go beyond the area. The purpose of setting this boundary is to assist participants in self-correction when they are in motion, preventing them from straying significantly from the camera's FOV. The camera can capture images from the real world up to 4K resolution. However, considering the balance between image quality and storage space, we use $640 \times 480$ as our implementary resolution, which means that low-cost RGB cameras with lower resolution are also applicable. For a better comparison, we follow previous studies~\cite{xue2021mmmesh} and extract the estimated rotation matrices of the unlocked joints (\ie, except the wrist, hand, ankle and foot joints) as the target of training while setting the rotation matrices of the rest joints as identity matrix.

\para{Experimental settings.}
We comprehensively evaluate the performance of \SystemName in various situations and compare it with open-source SOTA baselines. Unless otherwise specified, all experiments were carried out with a time-ordered split of 70\%, 20\%, and 10\% for the training, validation, and testing datasets, respectively. The Adam optimizer, with a default learning rate of 0.0003 and a batch size of 16, was adopted. The learning rate was decayed by a factor of 0.9 after each epoch, with a maximum of 50 training epochs and a patience of 5 epochs. Only the training dataset was segmented with a 1-frame overlap in temporal order, while the validation and test datasets were segmented without overlap.

\subsection{Data Collection}
\label{subsec:datacollection}
\begin{figure*}
  \centering
  \includegraphics[width=0.98\linewidth]{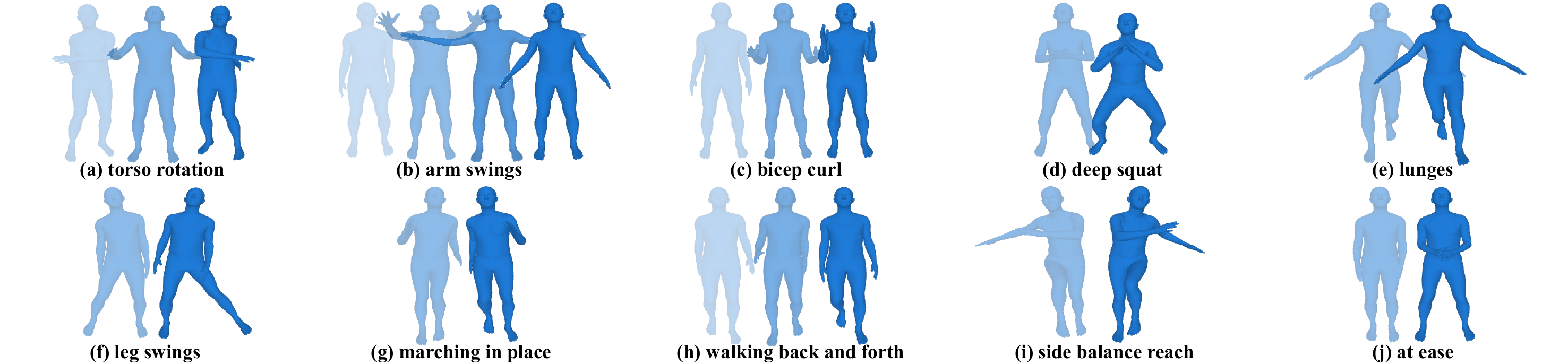}
  \caption{(a)--(c) are upper-limb activities; (d)--(f) are lower-limb activities; (g)--(j) are whole-body activities.}
  \label{fig:activities}
\end{figure*}
In this paper, we invited 16 participants (9 males and 7 females, aged 21 to 32, with heights ranging from \SI{1.58}{} to \SI{1.90}{\meter} and weights from \SI{49}{} to \SI{83}{\kilo\gram}) to perform 10 daily activities\footnote{This study has received the ethical approval from the authors’ institution.}. To comprehensively assess the perception capabilities of \SystemName across different body regions (\ie, upper limb, lower limb, whole body), as Fig.~\ref{fig:activities} shows, the 10 activities include: (a) torso rotation; (b) arm swings; (c) bicep curl; (d) deep squat; (e) lunges; (f) leg swings; (g) marching in place; (h) walking back and forth; (i) side balance reach; (j) at ease. For each activity, each participant continues to perform it for \SI{2}{\minute}. The sampling rates of mmWave radars and the RGB camera are \SI{10}{\hertz} and \SI{30}{\hertz}, respectively. As a result, the dataset contains \SI{12000}{} frames per participant, and the entire dataset contains more than \textbf{\SI{200000}{}} image-mmWave frames, including the data for the evaluation of micro-benchmarks.

\subsection{Evaluation Metrics}
\label{subsec:metrics}
We use the following metrics to evaluate the meshes reconstructed by our system for the activities mentioned above:
\begin{itemize}
    \item \textbf{Average Vertex Error (V)}~\cite{bogo2016keep, zhao2018rf}. The average vertex error by averaging the Euclidean distance between the vertices of reconstructed meshes and that of the corresponding GT\footnote{GT means the labels generated from the ground-truth RGB images.}.
    \item \textbf{Average Joint Localization Error (S)}~\cite{zhao2018rf, jiang2020towards}. The average skeleton error by averaging the Euclidean distance between the joint locations of reconstructed skeletons and the corresponding GT.
    \item \textbf{Average Joint Rotation Error (Q)}~\cite{xue2021mmmesh}. The average joint rotation error between the predicted joint rotations and the corresponding GT.
\end{itemize}
Note that we did not include some metrics (\eg, mesh location error, gender prediction accuracy) used in previous studies~\cite{xue2021mmmesh, xue2022m4esh, xue2023towards}. The reason is that \SystemName is a wearable system designed for self-sensing rather than sensing others. In such scenarios, users can provide accurate gender information and can use the body shape parameters (\ie, betas) estimated by the monocular camera directly. Furthermore, given that \SystemName is attached to a head-mounted host device, it shares the same coordinate system with the user. Therefore, \SystemName does not need to predict the global location and orientation of the wearer, and these metrics are not included.

\subsection{Overall Performance}
\label{subsec:overall}
\begin{figure*}
    \centering
    \begin{minipage}{0.25\linewidth}
    \centering
        \includegraphics[width=\linewidth]{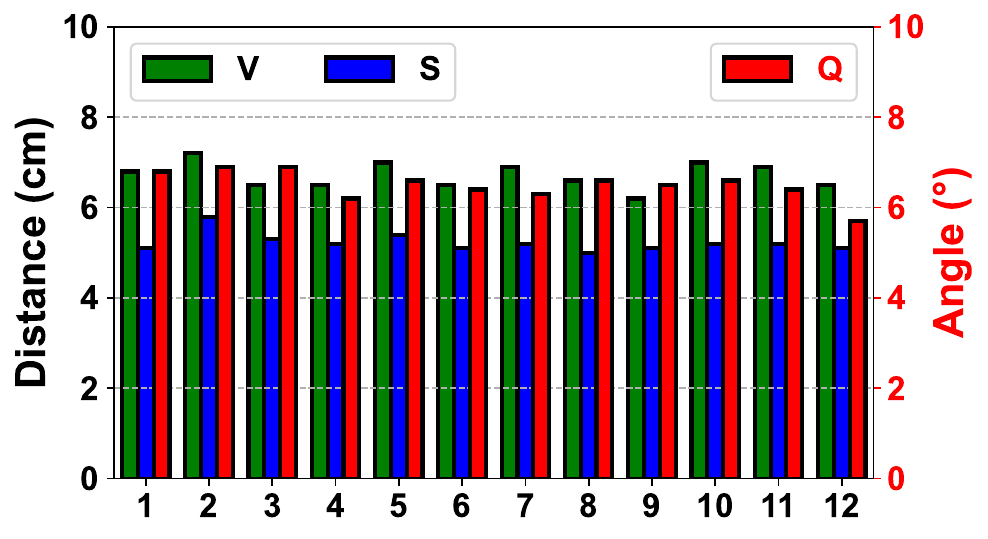}
        \vspace{-7mm}
        \caption{Overall.}
        \label{fig:overall_result}
        \vspace{-4mm}
    \end{minipage}
    \hspace{-2mm}
    \begin{minipage}{0.25\linewidth}
    \centering
        \includegraphics[width=\linewidth]{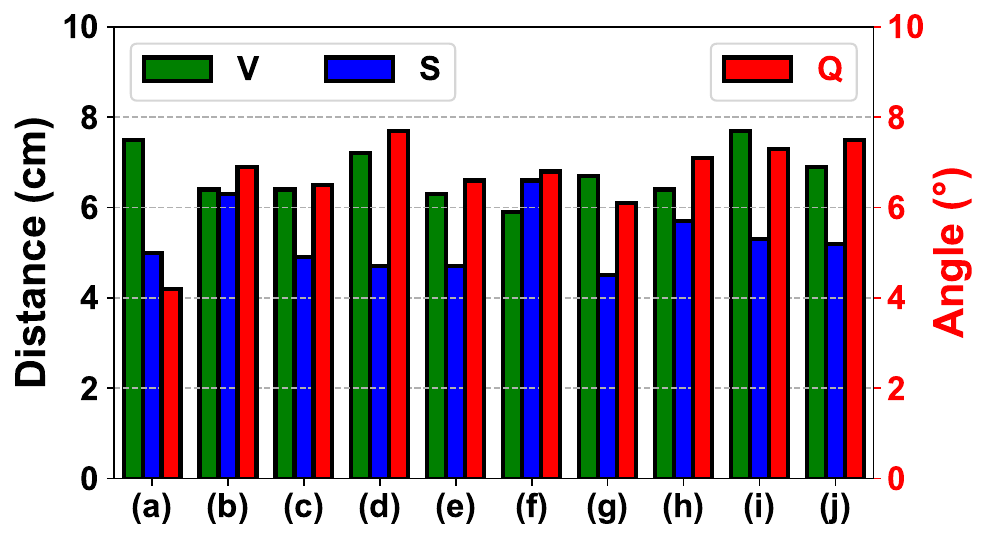}
        \vspace{-7mm}
        \caption{Different activities.}
        \label{fig:activities_result}
        \vspace{-4mm}
    \end{minipage}
    \hspace{-2mm}
    \begin{minipage}{0.25\linewidth}
    \centering
        \includegraphics[width=\linewidth]{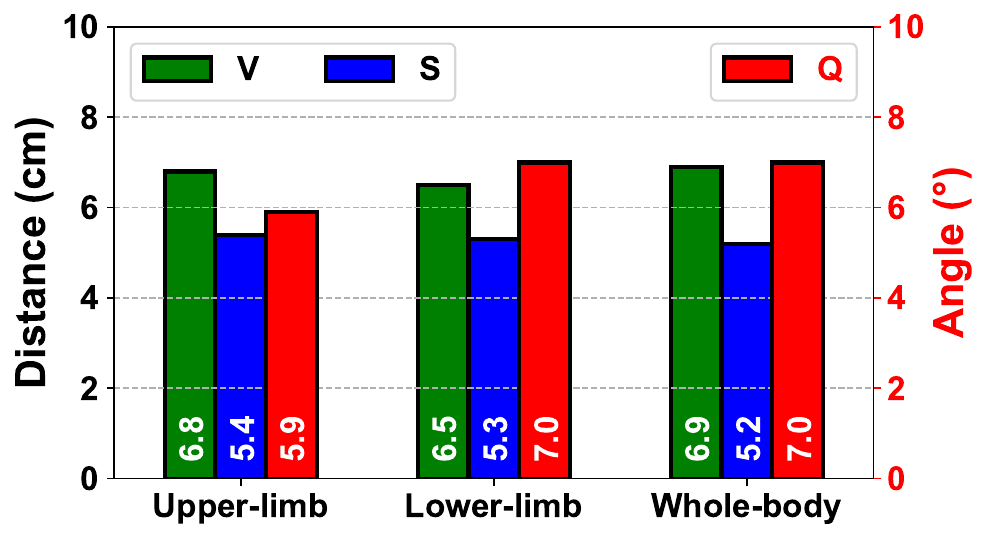}
        \vspace{-7mm}
        \caption{Different regions.}
        \label{fig:regions_result}
        \vspace{-4mm}
    \end{minipage}
    \hspace{-2mm}
    \begin{minipage}{0.25\linewidth}
    \centering
        \includegraphics[width=\linewidth]{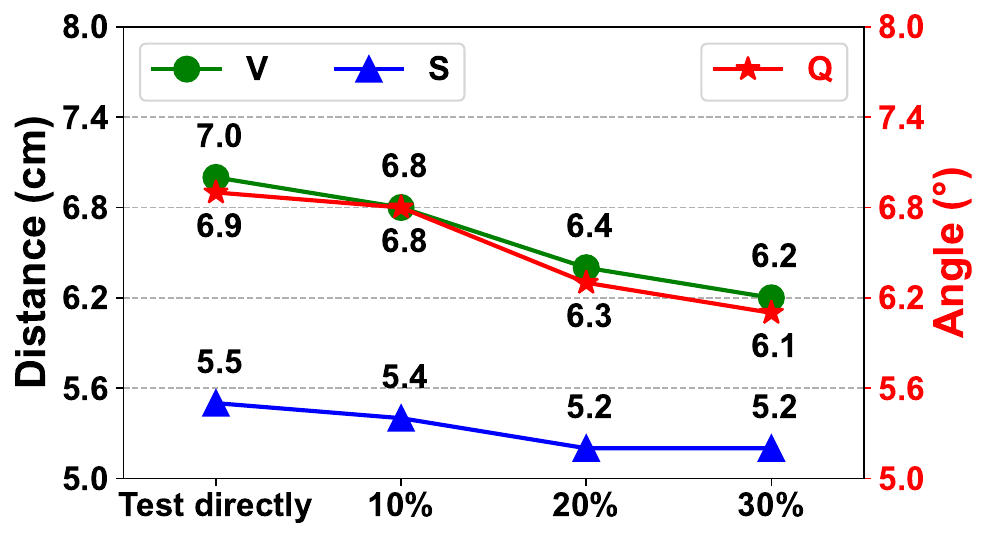}
        \vspace{-7mm}
        \caption{Unseen users.}
        \label{fig:unseen_result}
        \vspace{-4mm}
    \end{minipage}
\end{figure*}
We first exclude the data of four randomly selected users to serve as unseen users for further evaluation; their data are not included in any training or testing process except in Sec.~\ref{subsec:unseen}. Subsequently, we train a user-specific model for each user and use the set of these training data to train a basic model to evaluate performance on unseen users. As Fig.~\ref{fig:overall_result} shows, the performance of \SystemName across all participants is stable, with average V, S, and Q errors of \SI{6.8}{\centi\meter}, \SI{5.3}{\centi\meter}, and \SI{6.6}{\degree}, respectively. However, the errors for Participant 2 are the highest, which can be attributed to the participant's height of \SI{1.90}{\meter}. Following the signal processing described in Sec.~\ref{subsec:signal_processing}, the mmWave features over \SI{1.5}{\meter} will be eliminated by the Range-Gating; it does not adequately cover the lower body range. Furthermore, the absolute errors of the reconstructed meshes are positively correlated with the user's height.

We also conduct an in-depth analysis to investigate the performance of \SystemName in different activities and regions. The average errors for each activity are shown in Fig.~\ref{fig:activities_result}; we find that the performance varies significantly. In general, the errors are higher for actions with greater complexity. For example, the side balance reach activity presents the highest V error, averaging \SI{7.7}{\centi\meter}. The performance of \SystemName on the three different evaluation metrics does not show a significant correlation among activities. Although V error is as large as \SI{7.5}{\centi\meter} during torso rotation activity, Q error is only \SI{4.2}{\degree}. For different regions, the V and S errors are relatively stable, while the Q error for lower-limb and whole-body activities is \SI{1.1}{\degree} higher than that for upper-limb activities.

\subsection{Baselines}
\label{subsec:baseline}
\begin{table}
\centering
\caption{Comparison of \SystemName with SOTA baselines.}
\resizebox{0.95\linewidth}{!}{%
\begin{tabular}{l|c|c|c}
\toprule
\textbf{Baseline} & \textbf{V (cm)} & \textbf{S (cm)} & \textbf{Q ($^\circ$)} \\
\midrule
\begin{tabular}[c]{@{}l@{}}mmMesh\end{tabular} & \begin{tabular}[c]{@{}c@{}}\colorbar{9.4}{9.4}\end{tabular} & \begin{tabular}[c]{@{}c@{}}\colorbar{8.7}{8.7}\end{tabular} & \begin{tabular}[c]{@{}c@{}}\colorbar{9.9}{9.9}\end{tabular} \\
\hline
\begin{tabular}[c]{@{}l@{}}mmEgo\end{tabular} & \begin{tabular}[c]{@{}c@{}}\colorbar{7.9}{7.9}\end{tabular} & \begin{tabular}[c]{@{}c@{}}\colorbar{6.8}{6.8}\end{tabular} & \begin{tabular}[c]{@{}c@{}}\colorbar{7.5}{7.5}\end{tabular} \\
\hline
\begin{tabular}[c]{@{}l@{}}\textbf{Argus} (MLP)\end{tabular} & \begin{tabular}[c]{@{}c@{}}\colorbar{6.9}{6.9}\end{tabular} & \begin{tabular}[c]{@{}c@{}}\colorbar{5.3}{5.3}\end{tabular} & \begin{tabular}[c]{@{}c@{}}\colorbar{6.6}{6.6}\end{tabular} \\
\hline
\begin{tabular}[c]{@{}l@{}}\textbf{Argus} (KAN)\end{tabular} & \begin{tabular}[c]{@{}c@{}}\colorbar{6.5}{6.5}\end{tabular} & \begin{tabular}[c]{@{}c@{}}\colorbar{5.0}{5.0}\end{tabular} & \begin{tabular}[c]{@{}c@{}}\colorbar{6.4}{6.4}\end{tabular} \\
\bottomrule
\end{tabular}
}
\label{tab:comparison_result}
\end{table}
We compare \SystemName with two egocentric SOTA baselines~\cite{xue2021mmmesh, li2023egocentric} described in a recent study~\cite{li2023egocentric}; the implementation of these baselines adopts the open-source code~\cite{mmMeshGit, mmEgoGit}. Furthermore, to assess the effectiveness of KAN, we replace all KAN modules with MLP modules (with the same dimensions) as the third baseline. For fair comparison, we use data from the 12 participants in Sec.~\ref{fig:overall_result} to train models with three different neural networks, and test them on the testing data.

The results (Table~\ref{tab:comparison_result}) show that \SystemName outperforms the two baselines, achieving a 17.7--30.9\% reduction in V error and a 14.7--35.4\% reduction in Q error. Although the performance does not reach the levels achieved by a previous study~\cite{li2023egocentric}, it is important to note that the previous study was based on a high-capability, high-power, and high-cost radar, which features an equivalent $4 \times 4$ antenna array. In contrast, \SystemName utilizes stripped-down radars that feature an equivalent $2 \times 2$ antenna array. We overcome hardware limitations and, for the first time, achieve comparable performance with a system based on high-capability radars. By comparing the two versions of \SystemName, the effectiveness of KAN in handling such complex tasks involving high-dimensional data is evident.

\begin{figure*}
    \centering
    \begin{minipage}{0.2\linewidth}
    \centering
        \includegraphics[width=\linewidth]{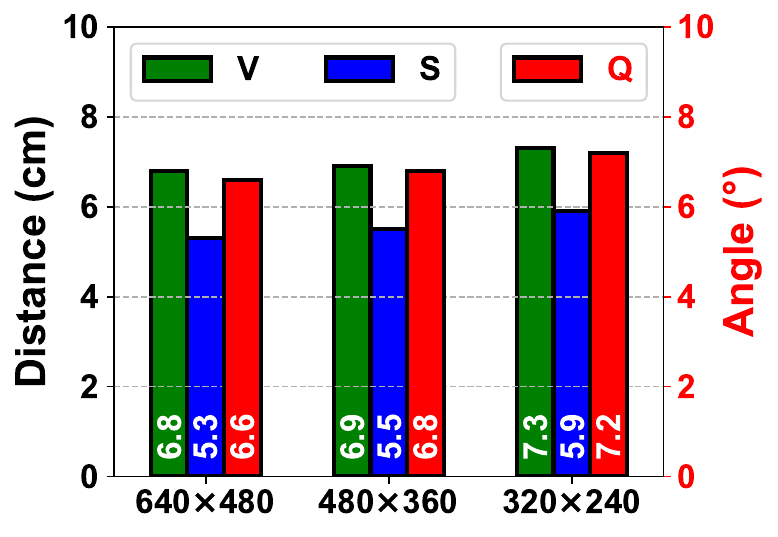}
        \vspace{-7mm}
        \caption{Resolutions.}
        \label{fig:resolution}
        \vspace{-4mm}
    \end{minipage}
    \hspace{-2mm}
    \begin{minipage}{0.2\linewidth}
    \centering
        \includegraphics[width=\linewidth]{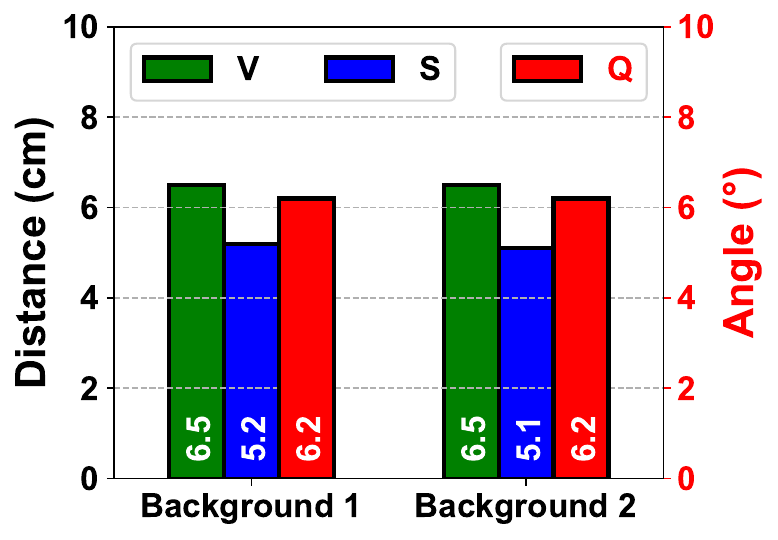}
        \vspace{-7mm}
        \caption{Backgrounds.}
        \label{fig:background}
        \vspace{-4mm}
    \end{minipage}
    \hspace{-2mm}
    \begin{minipage}{0.2\linewidth}
    \centering
        \includegraphics[width=\linewidth]{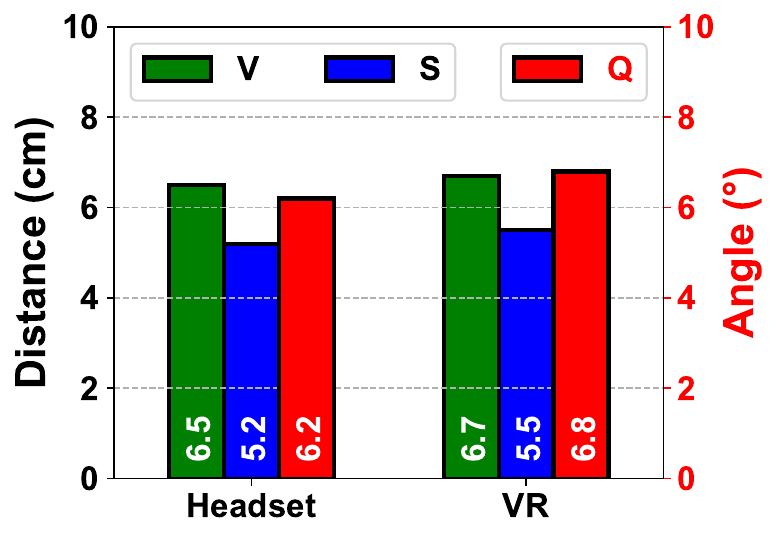}
        \vspace{-7mm}
        \caption{Host devices.}
        \label{fig:devices}
        \vspace{-4mm}
    \end{minipage}
    \hspace{-2mm}
    \begin{minipage}{0.2\linewidth}
    \centering
        \includegraphics[width=\linewidth]{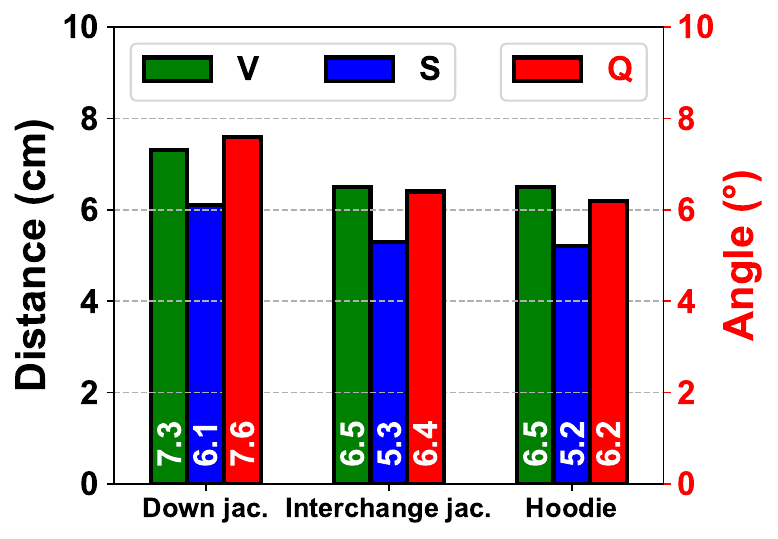}
        \vspace{-7mm}
        \caption{Clothes.}
        \label{fig:clothes}
        \vspace{-4mm}
    \end{minipage}
    \hspace{-2mm}
    \begin{minipage}{0.2\linewidth}
    \centering
        \includegraphics[width=\linewidth]{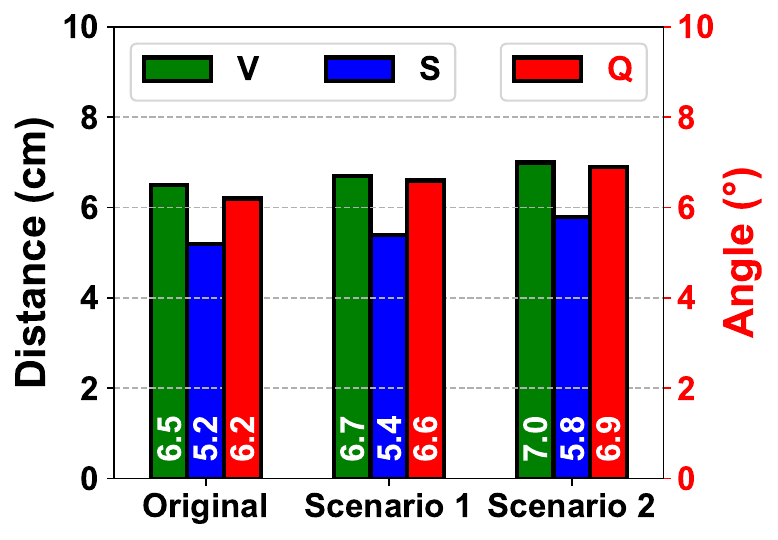}
        \vspace{-7mm}
        \caption{Multipath.}
        \label{fig:multipath}
        \vspace{-4mm}
    \end{minipage}
\end{figure*}
\subsection{Unseen User Performance}
\label{subsec:unseen}
Considering real-world application scenarios, we evaluate the performance of \SystemName on unseen users (2 males and 2 females). Specifically, we first evaluate the basic model's performance on them by testing directly (without fine-tuning). Then, we fine-tune the basic model using a sub-dataset that accounts for $x$\% of all training data for each participant. By increasing the value of $x$, our goal is to explore how much data is required to fine-tune the basic model to achieve satisfactory results for an unseen user. As Fig.~\ref{fig:unseen_result} shows, when the training data used for fine-tuning reaches 30\% of the total, the model achieves remarkable performance. Furthermore, owing to the generalization obtained from training data across multiple users, the performance of the model fine-tuned with 30\% of the new user data surpasses the average performance of the user-specific models. Therefore, when deploying \SystemName for a new user, theoretically only $14 \times 0.3=$ \SI{4.2}{\min} of the new data collection is needed to deploy.

\subsection{Other Micro-benchmarks}
\label{subsec:benchmarks}
We also evaluate several factors that may influence the performance of \SystemName and report them as micro-benchmarks.

\para{Image resolution.} Most current mobile devices can capture RGB images with resolutions higher than $640 \times 480$. To simulate budget devices with lower-quality cameras, we downsample the RGB images to $480 \times 360$ and $320 \times 240$, evaluating the performance under lower-quality labels.

\para{Image background.} To evaluate the impact of the experimental background on the effectiveness of labels extracted by the proposed MoCap system, we invite two participants to record a new set of data at different experimental sites.

\para{Host device.} We select two participants and replace the host device with a VR (\ie, HTC VIVE Pro Eye~\cite{htcvive}), using the same magnetic attachment method as shown in Fig.~\ref{subfig:detachable}. A new set of data is collected for each participant to evaluate the performance of \SystemName on different host devices.

\para{Type of clothing.} To evaluate the robustness of \SystemName, we select two participants and collect data while they wear different clothing. Specifically, for each participant, in addition to the data (wearing a hoodie) collected in Sec.~\ref{subsec:overall}, we also collect data while they wear an interchangeable jacket and a down jacket, with all other experimental settings the same.

\para{Multipath effect.}
Since mmWave sensing suffers from the multipath effect, we evaluated the robustness of \SystemName to scatters and reflectors, such as clothes and furniture. Two participants collected additional testing data with sundries placed close by without hindering activities. In Scenario 1, a chair with a backpack was placed next to the participant. In Scenario 2, an additional chair with a down jacket was added. Testing the model trained in a clutter-free scenario allowed us to assess the robustness to multipath effects.

The results are listed in Fig.~\ref{fig:resolution}--\ref{fig:multipath}, \SystemName is robust to image resolution and background. For different devices, the variation in geometry directly affects the deployment angle and spatial position of \SystemName, and the results show that deploying \SystemName on a VR system leads to V and S errors that are approximately 3\% and 5.7\% higher, respectively, compared to deploying it on a headset. The user's clothing significantly affects system performance. Clothing with large reflective surfaces, such as down jackets, exacerbates both self-occlusion and specular reflection issues. As a result, the inability to precisely estimate joint rotations and positions leads to substantial errors in the results.

\subsection{Computational Delay}
\label{subsec:delay}
To verify the practicality of \SystemName, we evaluate its computational delay. Unlike previous studies~\cite{xue2021mmmesh,li2023egocentric}, which used TI xWRxx43 radars with well-supported C-based signal processing to obtain point clouds, our work involved engineering efforts to process raw signals from BGT60TR13C radars, which lack such support. We developed a Python-based signal processing pipeline, allowing for greater customization of processing details. Despite the current signal processing time being \SI{0.880}{\second} on an Intel Core i7 CPU, the model inference and SMPL rendering times are \SI{11.6}{\milli\second} and \SI{1.5}{\milli\second}, respectively, leading to an overall algorithmic delay of \SI{0.89}{\second}.

Based on a previous study~\cite{xue2021mmmesh} using a AWR1843BOOST radar, which reported a \SI{28}{\milli\second} processing time for converting raw signals to point clouds, we reasonably estimate that transitioning our signal processing to the C programming language and leveraging hardware acceleration could reduce the delay to within \SI{80}{\milli\second} (approximately a 10x speedup from Python). \SystemName demonstrates the feasibility of egocentric HMR using stripped-down radars. If well-supported C-based signal processing and hardware acceleration are supported for the radar, real-time HMR with an algorithmic delay of less than \SI{100}{\milli\second} per frame is achievable.
\section{Limitations \& Future Work}
\label{sec:limitation_future}
While our study shows the effectiveness of \SystemName for egocentric HMR using stripped-down radars in a multi-view manner, there are limitations that need further investigation. Addressing them can help improve performance.

\para{Restricted activities.} The hardware design of \SystemName limits the detection of activities above the ear level, such as raising arms overhead, resulting in restricted activities in HMR. However, the supported sensing range covers most daily activities. To overcome this limitation, adding additional sensors, oriented differently, could be a promising solution to cover the sensing blind spots.

\para{More sophisticated hardware.} \SystemName is a prototype featuring a Raspberry Pi as an intermediary between mobile devices and radars. Future work could involve designing a dedicated, more compact, and faster PCB as the intermediary. Moreover, a more powerful radar model has been released recently (\ie, BGT60ATR24C~\cite{BGT60ATR24C}) with 2 transmit antennas and 4 receive antennas are available, which could further enhance the system’s performance in HMR task.
\section{Conclusion}
\label{sec:conclusion}
This paper introduces a novel mmWave-based add-on system named \SystemName, which is the first wearable add-on based on stripped-down mmWave radars deployed in a multi-view configuration for egocentric HMR. The key idea of \SystemName is that the compact, limited-capability mmWave radars on both the left and right sides can form a multi-view sensing system that mitigates self-occlusion and specular reflection issues in such an egocentric view through complementary viewpoints. By addressing three unique challenges, the limitations caused by the stripped-down radar are mitigated, achieving performance comparable with solutions based on high-capability radars. Extensive evaluation demonstrates the robustness and practicality of \SystemName, making it a promising alternative to existing HMR solutions.


\balance
\bibliographystyle{ACM-Reference-Format}
\bibliography{reference}










\end{document}